%% file: top.tex
\documentclass[sigconf, nonacm, pdfa]{acmart}

\newcommand\vldbdoi{10.14778/3681954.3682027}
\newcommand\vldbpages{3644 - 3656}
\newcommand\vldbvolume{17}
\newcommand\vldbissue{11}
\newcommand\vldbyear{2024}
\newcommand\vldbauthors{\authors}
\newcommand\vldbtitle{\shorttitle} 
\newcommand\vldbavailabilityurl{URL_TO_YOUR_ARTIFACTS}
\newcommand\vldbpagestyle{empty}

\usepackage{nicefrac}

\include{packages-commands}

\usepackage[a-2b]{pdfx}

\usepackage{float}
\usepackage{siunitx}

\begin{document}

\title{Aleph Filter: To Infinity in Constant Time}

\author{{Niv Dayan}}
\affiliation{
	\institution{University of Toronto}
	\country{Canada}
}
\email{nivdayan@cs.toronto.edu }

\author{Ioana-Oriana Bercea}
\affiliation{
	\institution{KTH Royal Institute of Technology}
	\country{Sweden}
}
\email{bercea@kth.se}

\author{Rasmus Pagh}
\affiliation{
	\institution{BARC, University of Copenhagen}
	\country{Denmark}
}
\email{pagh@di.ku.dk }

\begin{abstract}

Filter data structures are widely used in various areas of computer science to answer approximate set-membership queries.  In many applications, the data grows dynamically, requiring their filters to expand along with the data. However, existing methods for expanding filters cannot maintain stable performance, memory footprint, and false positive rate (FPR) simultaneously. We address this problem with Aleph Filter, which makes the following contributions. (1) It supports all operations  (insertions, queries, deletes, etc.) in constant time, no matter how much the data grows. (2) Given an estimate of how much the data will ultimately grow, Aleph Filter provides a memory vs. FPR trade-off on par with static filters.

\end{abstract}

 \maketitle
 
 \pagestyle{\vldbpagestyle}
 \begingroup\small\noindent\raggedright\textbf{PVLDB Reference Format:}\\
 \vldbauthors. \vldbtitle. PVLDB, \vldbvolume(\vldbissue): \vldbpages, \vldbyear.\\
 \href{https://doi.org/\vldbdoi}{doi:\vldbdoi}
 \endgroup
 \begingroup
 \renewcommand\thefootnote{}\footnote{\noindent
 	This work is licensed under the Creative Commons BY-NC-ND 4.0 International License. Visit \url{https://creativecommons.org/licenses/by-nc-nd/4.0/} to view a copy of this license. For any use beyond those covered by this license, obtain permission by emailing \href{mailto:info@vldb.org}{info@vldb.org}. Copyright is held by the owner/author(s). Publication rights licensed to the VLDB Endowment. \\
 	\raggedright Proceedings of the VLDB Endowment, Vol. \vldbvolume, No. \vldbissue\ %
 	ISSN 2150-8097. \\
 	\href{https://doi.org/\vldbdoi}{doi:\vldbdoi} \\
 }\addtocounter{footnote}{-1}\endgroup
 
 \ifdefempty{\vldbavailabilityurl}{}{
 	\vspace{.3cm}
 	\begingroup\small\noindent\raggedright\textbf{PVLDB Artifact Availability:}\\
 	The source code, data, and/or other artifacts have been made available at  \url{https://github.com/nivdayan/AlephFilter}.
 	\endgroup
 }
 
 

\input{introduction}

\input{background}

\input{problem}

\input{main}

\input{evaluation}

\input{related-work}

\input{conclusion}


\begin{acks}
We thank the  reviewers for their valuable feedback. This research was supported NSERC grant \#RGPIN-2023-03580. 
\end{acks}

\bibliographystyle{ACM-Reference-Format}
\balance
{
\bibliography{library-short}
}
\appendix

\end{document}

%% file: packages-commands.tex
\usepackage{array}
\usepackage{wrapfig}

\usepackage{graphicx}
\usepackage{url}
\usepackage{balance}

\usepackage{mathtools}

\usepackage[linesnumbered,noend]{algorithm2e}
\usepackage{comment}

\usepackage{enumitem}

\newcommand{\eat}[1]{}
\newcommand{\Paragraph}[1]{\vspace{-0.02in}\smallskip\noindent{\bf #1.}}

\usepackage{fixme}
\fxsetup{
    status=draft,
    author={Note},
    layout=inline,
    theme=color,
    silent=true,
    inline=true
}
\FXRegisterAuthor{ma}{ama}{\color{red}Manos}
\FXRegisterAuthor{si}{sia}{\color{blue}Stratos}
\FXRegisterAuthor{nd}{nda}{\color{magenta}Nov}

%% file: introduction.tex
\section{Introduction} \label{sec:intro}

\Paragraph{Filters} A filter is a compact data structure that represents keys in a set and answers set-membership queries \cite{Pandey2024}. It cannot return a false negative, but it returns a false positive with a probability that depends on the amount of memory used on average to represent each key. A filter is typically stored at a higher layer of the memory hierarchy (e.g., DRAM) than the data keys that it represents, which typically reside in storage (i.e., disk or SSD) or over a network. If a filter returns a negative, the sought-after key is guaranteed not to exist, meaning the full data set does not need to be searched. Thus, filters eliminate storage accesses and/or network hops to improve a system's performance \cite{Dayan2017, Dayan2018, Dayan2019, Sarkar2023, Idreos2019}.

\Paragraph{The Need for Dynamic Filters} Applications with dynamic data require filters that support deletes and can expand. 
Various modern key-value stores employ dynamic filters to map data entries in storage \cite{Chandramouli2018, Andersen2009, Debnath2010, Debnath2011, Dayan2021, Dayan2021B, Ren2017, Conway2023, Wang2024grf}, while network applications use them to support black lists and multicast routing \cite{Wu2021}.

Bloom filter \cite{Bloom1970, Tarkoma2012, Broder2002}, the oldest and best-known filter, does not support deletes or efficient expansion. If the data grows or changes, the only recourse is to recreate its Bloom filter from scratch by rereading the original keys. 
This, however, can be performance-prohibitive if the keys reside in storage. In contrast, dynamic filters support deletes and expansion \textit{without} rereading the original keys. 

\Paragraph{Motivation: Scaling Dynamic Filters} As existing dynamic filters expand, their performance,  memory footprint, and/or  false positive rate (FPR) deteriorate. Our research goal is to better scale these cost metrics so that dynamic filters remain effective no matter how much the data grows or changes. 

\Paragraph{Quotient Filters} 
Quotient filters are a family of dynamic filters that store a fingerprint for each key in a compact hash table \cite{Carter1978} and handle collisions using linear probing \cite{Clerry1984, Pandey2017, Dillinger2009, Bender2012, Pandey2021, Reviriego2024}. 
While they seamlessly support deletions, expanding them efficiently is more challenging. 
The reason is that we do not have access to the original keys and can therefore not rehash each key to a unique slot within a larger hash table. Existing quotient filters address this issue by transferring one bit from each entry's fingerprint to become a part of its slot address to evenly distribute the fingerprints across a 2x larger  hash table \cite{Bender2012, Zhang2021}. As a result, the fingerprints shrink as the data grows,  causing the FPR to increase rapidly. Eventually, the fingerprints run out of bits, making the filter useless by returning a positive for any query. 

\Paragraph{The State of the Art: InfiniFilter} The recent InfiniFilter \cite{Dayan2023}  is a quotient filter that can expand indefinitely while better scaling the FPR. It does so using a hash slot format that supports variable-length fingerprints. Thus, while fingerprints of older entries shrink across expansions, the fingerprints of new entries can still be initialized to occupy the original slot length. This design keeps fingerprints long on average. As a result, the FPR increases more slowly. Nevertheless, InfiniFilter exhibits two remaining scalability challenges. 

\Paragraph{Problem 1: Performance Scalability}  After the first few expansions, the fingerprints of older entries within InfiniFilter run out of bits. Such entries are referred to as void entries \cite{Dayan2023}. A void entry cannot be mapped to a unique slot in a 2x larger hash table since there is no remaining fingerprint bit to transfer to its slot address.
InfiniFilter tackles this problem by transferring and storing void entries along a series of smaller hash tables. However, this causes queries and deletions to potentially search multiple hash tables,  increasing their CPU overheads. Can we better scale the costs of queries and deletions? 

\Paragraph{Problem 2: FPR vs. Memory Scalability} As InfiniFilter expands, the shorter fingerprints of older entries cause the FPR to increase. To counteract this, InfiniFilter can assign even longer fingerprints to newer entries as we expand to cause the FPR to converge \cite{Dayan2023}. Nevertheless, this entails widening the slot width and thus inflating the memory footprint. Thus, there is an intrinsic scalability contention between the FPR and memory for InfiniFilter and for expandable filters in general \cite{Pagh2013}. Is it possible to alleviate this contention so that the FPR vs. memory trade-off resembles that of static filters? 

\Paragraph{Aleph Filter} We propose Aleph Filter\footnote{The name Aleph is borrowed from set theory, where the Aleph numbers denote different orders of infinity. }, an infinitely expandable filter with constant time performance for all operations and superior memory vs. FPR trade-offs. 
It builds on InfiniFilter in three ways. 

\Paragraph{Contribution 1: Faster Queries by Duplicating Void Entries} 
During expansion, Aleph Filter duplicates every void entry in the expanded hash table across both slots that it could have been mapped to if there had been an additional fingerprint bit to sacrifice. This keeps the information about all entries, including void entries, within one main hash table. As a result, Aleph filter accesses only one hash table for any query in $O(1)$ time. We show how to tune the filter so that duplicated void entries occupy negligible space. 

\Paragraph{Contribution 2: Faster Deletes using Tombstones} 
Duplicating void entries within the main hash table complicates deletions as potentially multiple duplicates must be identified and removed when an older entry is deleted. Aleph filter addresses this challenge by first transforming the target void entry into a tombstone. Before the next expansion, it retrieves the original hash of the deleted void entry and uses it to identify and remove all of its duplicates. The overhead of removing these duplicates is negligible and gets amortized as a part of the next expansion. 

\Paragraph{Contribution 3: Inverting the FPR vs Memory Trade-Off} In many applications, the maximum data size can be predicted in advance. We show that given a reasonable estimate of how much the data will grow, we can pre-allocate slightly longer fingerprints from the onset and assign shorter fingerprints as we expand. When we reach the estimated data size, the filter guarantees an FPR vs. memory  trade-off that is on par with that of static filters. 

%% file: background.tex
	\renewcommand{\arraystretch}{1.0}
	\begin{table}[t]
				\caption{ Terms used throughout the paper to describe Aleph Filter and other baselines.     } \label{tab:terms}
		\begin{tabular}{ |p{0.6cm}|p{7cm}|}  
			\hline
			term & definition \\
			\hline
			$N$ 	  		 					  &    current filter capacity divided by initial capacity 	    \\
			$X$ 	  		 					  &   number of expansions so far (i.e., $X = \lceil \log_2(N) \rceil$) 	    \\
			$M$ 	  		 					  &    total memory used for the filter (bits / entry)  	    \\
			$F$ 	  		 					  &    initial fingerprint length when the filter is first allocated  \\
			$h(...)$ 	  		 					  &    mother hash generating function   \\
			$\alpha$ 	  		 		&   fraction of occupied slots  ($0 \le \alpha < 1$)    \\
			\hline
		\end{tabular}
	\end{table}
	
\section{Background } \label{sec:background_sec}

This section describes Quotient Filter and InfiniFilter, on top of which we build Aleph Filter. 

\subsection{Quotient Filter}

A quotient filter \cite{Clerry1984, Dillinger2009, Bender2012} is a hash table that stores a fingerprint for each inserted key. It  does this by first generating a \textit{mother hash} for a key consisting of $F + A$ bits using some hash function $h(...)$. The least significant $A$ bits of this mother hash represent the key's \textit{canonical slot}. The subsequent $F$ bits are the key's fingerprint. Table~\ref{tab:terms} lists terms used throughout the paper. 

A quotient filter resolves hash collisions using Robin Hood hashing \cite{Celis1985}, which is a variant of linear probing \cite{Peterson1957}. This means that all fingerprints mapped to a given canonical slot are stored contiguously, and they push to the right any other existing fingerprints that they collide with. A \textit{run} is defined as a group of contiguous fingerprints belonging to the same canonical slot. A \textit{cluster} is defined as several adjacent runs where all but the first have all been pushed to the right due to collisions. 

\begin{figure}[t]
	\centering
	\includegraphics[width=8.5cm]{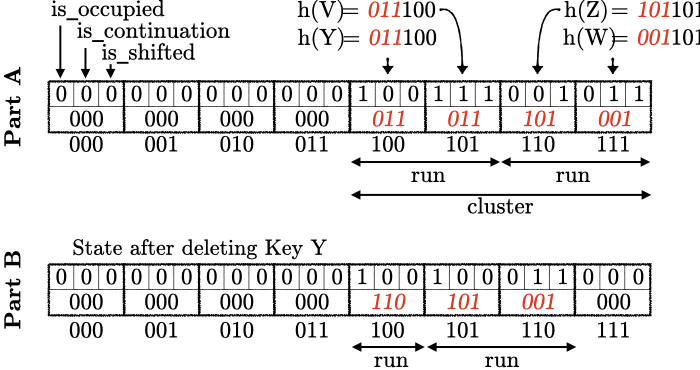}
	\vspace{-6mm}
	\caption{\small  A Quotient filter stores a fingerprint for each key in a hash table, and it resolves has collisions by organizing fingerprints into runs and clusters. Fingerprints are illustrated in {\color{black} italicized} red.}
	\vspace{-4mm}
	\label{fig:quotient_filter}
\end{figure}

Figure \ref{fig:quotient_filter} Part A illustrates a quotient filter with eight slots after four insertions  {\color{black}in any order} of Keys $V$, $Y$, $Z$ and $W$. The mother hash for each key is shown at the top. The rightmost (least significant) bits for each mother hash represent the key's canonical address. The subsequent bits, shown in {\color{black} italicized} red, are the fingerprint. 
Keys $V$ and $Y$ share Canonical Slot 100 while Keys $Z$ and $W$ share the adjacent Canonical Slot 101. The result  is a cluster consisting of two runs, each with two slots. 

\Paragraph{Metadata Bits} To keep track of the start and end of runs and clusters, a quotient filter employs three metadata bits per slot. The is$\_$occupied bit indicates whether the slot is a canonical slot for at least one existing key. The is$\_$shifted bit indicates whether the slot contains a fingerprint that has been shifted to the right relative to its canonical slot. The is$\_$continuation bit indicates whether the  slot contains the start of a new run within a cluster. 

In Figure \ref{fig:quotient_filter} Part A, the is$\_$occupied flag is set for Slots 100 and 101 as each of them is a canonical slot for at least one key. The is$\_$shifted flag is set for Slots 101, 110, and 111 as they each contain a fingerprint that has been shifted to the right from its canonical slot. The is$\_$continuation flag is set for Slots 101 and 111 as they each contain a fingerprint belonging to a run that starts to its left.

{ \color{black}
\Paragraph{Query} To illustrate the query process, consider a query to entry $W$ in figure \ref{fig:quotient_filter} Part (A). The query begins at the canonical slot for the target key (Slot 101). If the is$\_$occupied flag for this slot is set to 0, the key could not have been inserted so the query returns a negative. 
In this case, however, the is$\_$occupied flag is set to 1 and so the search continues. 
The query must  now find the start of the target run that might contains the sought-after key, yet this run could have been pushed to the right from its canonical slot due to collisions. 
To find the target run, the query first moves leftwards until reaching the start of the cluster. 
Along the way, it counts the number of  is$\_$occupied flags $c$ set to ones (in our example $c=2$), each of which indicates the existence of a run in the cluster that precedes Run $r$. After reaching the start of the cluster (at Slot 100), the query moves rightwards while skipping $c$ runs, at which point it reaches the target run (at Slot 110). It then scans this run, returning a positive if it finds a matching fingerprint (at Slot 111). 
}

\begin{figure}[t]
	\vspace{-0mm}
	\includegraphics[width=8.9cm]{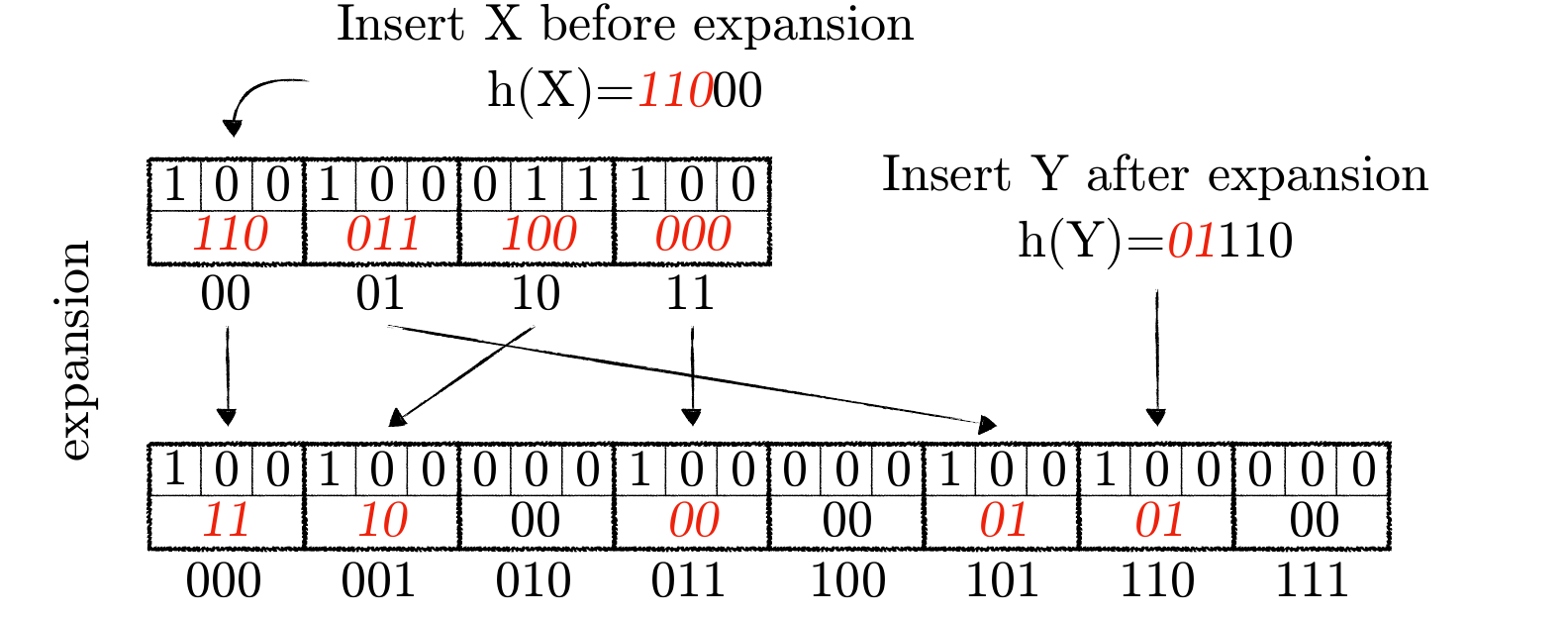}
	\vspace{-5mm}
	\caption{ The Fingerprint Sacrifice method transfers one bit from the fingerprint to the slot address while expanding to evenly distribute the entries across the larger hash table.     }
	\vspace{-3mm}
	\label{fig:quotient_filter_expansion}
\end{figure}

\Paragraph{Insertion} An insertion commences  as a query by first finding the start of the target run for the key we wish to insert. It then adds a fingerprint to this run, pushing subsequent fingerprints in the cluster one slot to the right to clear space. This may cause more runs to join the cluster by pushing them  from their canonical slots. 

\Paragraph{Delete} A delete operation also commences like a query by first finding the target  run of the key we wish to remove. It then scans the run and removes the first matching fingerprint. Then, it pulls all remaining fingerprints in the cluster one slot to the left to keep the cluster contiguous. For example, Figure \ref{fig:quotient_filter} Part B illustrates how deleting Key~$Y$ pulls the second run in the cluster back to its canonical slot, causing the cluster to fragment. 
As with any tabular filter, it is only permissible to delete keys that we know for sure had been inserted to prevent false negatives. 

\Paragraph{False Positive Rate} A quotient filter's FPR is \mbox{$\approx\alpha \cdot 2^{-F}$}, where $F$ is the fingerprint size and $\alpha$ is the fraction of the slots that are occupied. The intuition is that the target run contains on average $\alpha$ fingerprints, while the probability that each of them  matches that of the sought-after key is $2^{-F}$. 


\Paragraph{Reaching Capacity} As long as the quotient filter is less than $90\%$ full (i.e., $\alpha=0.9$), the clusters stay small on average leading to constant time operations. As it exceeds $90\%$ utilization, however, the clusters' lengths begin to grow rapidly. This causes performance to plummet as all operations on the filter must traverse a greater number of slots. To keep performance stable and to accommodate more insertions, it is desirable to expand the filter. 



\Paragraph{Fingerprint Sacrifice} The standard approach for expanding a quotient filter is to derive the mother for every key by concatenating its canonical slot address to its fingerprint.  We then reinsert the mother hash  into a hash table with twice the capacity of the original one \cite{Bender2012}. 
Figure~\ref{fig:quotient_filter_expansion} illustrates an example.
As shown, this approach  transforms the least significant bit of each fingerprint to the most significant bit of its canonical slot address in order to to evenly distribute the fingerprints across the larger hash table.  
Hence, all fingerprints shrink by one bit in each expansion, causing the FPR to double, i.e., the FPR becomes $O(N \cdot 2^{-F})$ \cite{Dayan2023}. This method supports at most $F$ expansions, at which point all fingerprints run out of bits. We summarize the properties of this method in Row 1 of Table \ref{tab:table1}. 

\begin{figure}[t]
	\includegraphics[width=9cm]{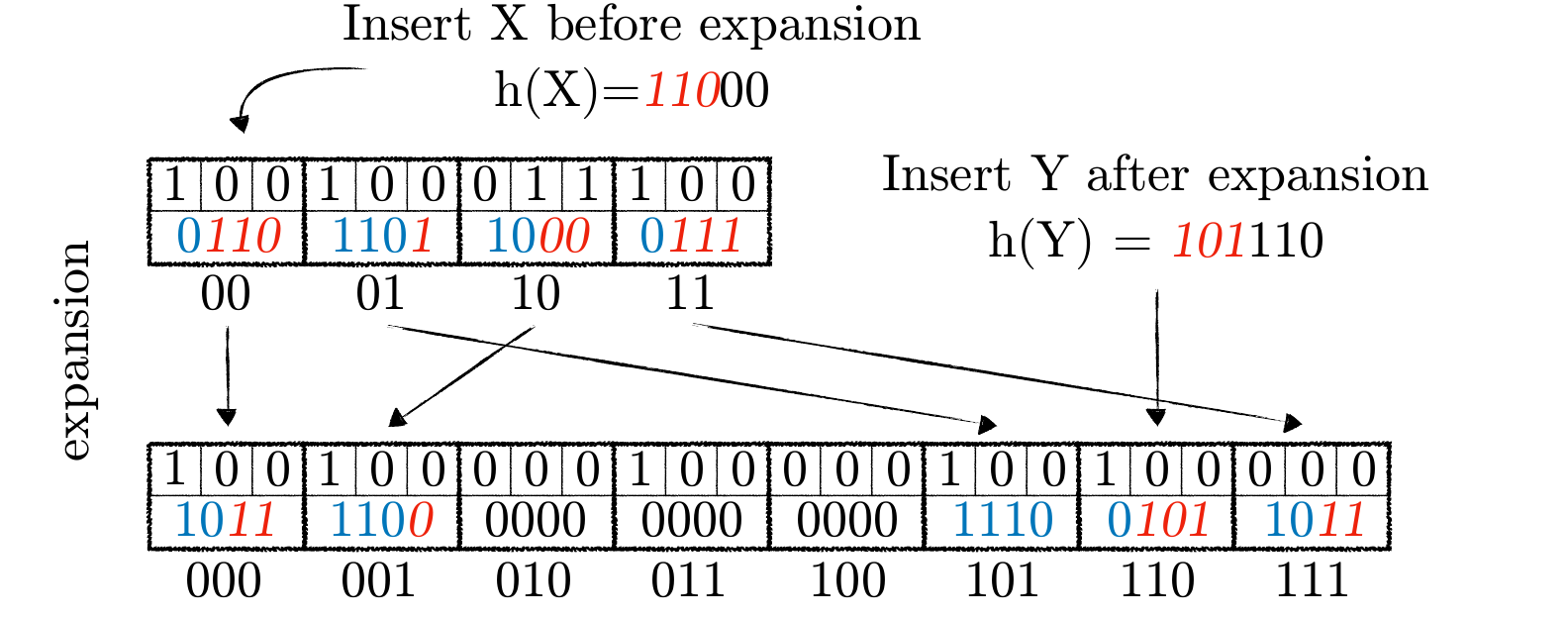}
	\vspace{-5mm}
	\caption{ By supporting variable-length fingerprints, InfiniFilter can set long fingerprints to newer entries inserted after expanding.   }
	\vspace{-0mm}
	\label{fig:infinifilter}
\end{figure}

\subsection{InfiniFilter}

InfiniFilter is a quotient filter that sets longer fingerprints to newer entries to allow expanding indefinitely while better scaling the FPR. 

\Paragraph{Unary Padding} The central  innovation of InfiniFilter is the ability to store variable-length fingerprints. This is achieved by padding each fingerprint with a self-delimiting unary code to occupy the rest of the space in the slot. The top part of Figure \ref{fig:infinifilter} illustrates an instance of InfiniFilter with four occupied slots and fingerprints of lengths 3, 1, 2 and 3 bits from left to right. Unary codes and fingerprints and  are shown in blue and {  \color{black} italicized} red, respectively. 

\Paragraph{Expansion Algorithm} As with the Fingerprint Sacrifice method, InfiniFilter  transfers the least significant bit from each fingerprint to become the most significant bit of the slot address during expansion to evenly distribute existing entries across the expanded hash table. After the expansion, InfiniFilter's slot format allows inserting longer fingerprints than the older ones that had shrunk. This is shown by the insertion of entry $Y$ in Figure \ref{fig:infinifilter}. This entry is assigned a three bit fingerprint despite the fact that fingerprints that existed before the expansion have now all shrunk to two or fewer bits. This keeps the average fingerprint length longer.

\begin{figure}[t]
	\vspace{-2mm}
	\includegraphics[width=9cm]{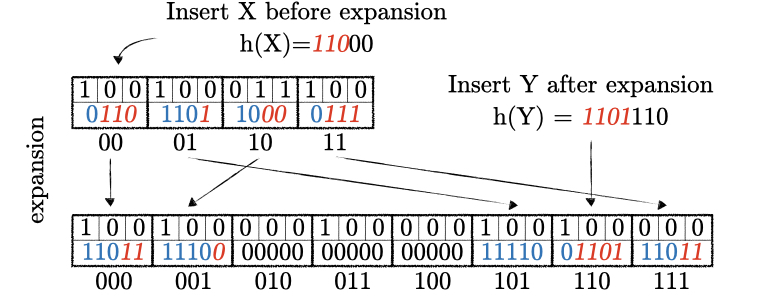}
	\vspace{-6mm}
	\caption{ InfiniFilter in the Widening Regime gradually  increases the slot width across expansions to keep the false positive rate constant.    }
	\vspace{-5mm}
	\label{fig:infinifilter_growing}
\end{figure}

\Paragraph{False Positive Rate} Since InfiniFilter doubles in capacity every time it expands, the fingerprints within the filter follow a geometric distribution with respect to their lengths: half of them are as long as possible, a quarter are shorter by one bit, an eighth are shorter by two bits, etc. Generally, the fingerprints created $i$ generations ago comprise a fraction of $\approx \alpha \cdot 2^{-i-1}$ of the occupied slots, while their fingerprints have a length of $F-i$ bits and hence a collision probability of $2^{-F+i}$.  The  FPR after $X$ expansions can be derived as a weighted average of these terms: $\sum_{i=0}^{X} 2^{-F+i} \cdot \alpha \cdot 2^{-i-1} \lesssim (\log_2(N)  +2) \cdot 2^{-F-1} \cdot \alpha$.  After $\approx 2^{F}$ expansions, the FPR reaches one, at which point the filter returns a positive for any query.




\Paragraph{Fixed-Length vs. Widening Regimes} To further scale the FPR, it is possible to widen slots across expansions. For example, Figure \ref{fig:infinifilter_growing} illustrates increasing the slot width by one bit. Entries inserted after the expansion are set fingerprints of length four rather than three bits. By assigning fingerprints of length  \mbox{$F + \lceil 2 \cdot \log_2(X+1) \rceil$} bits to new entries after the $X^{\text{th}}$ expansion, the FPR converges to a constant smaller than $2^{-F}$ \cite{Dayan2023}. The intuition is that the longer fingerprints of newer entries make their weighted contribution to the FPR vanishingly small. 
We refer to this as the Widening Regime, in contrast to the Fixed-Width Regime described just before.

\begin{figure}[t]
	\includegraphics[width=6.5cm]{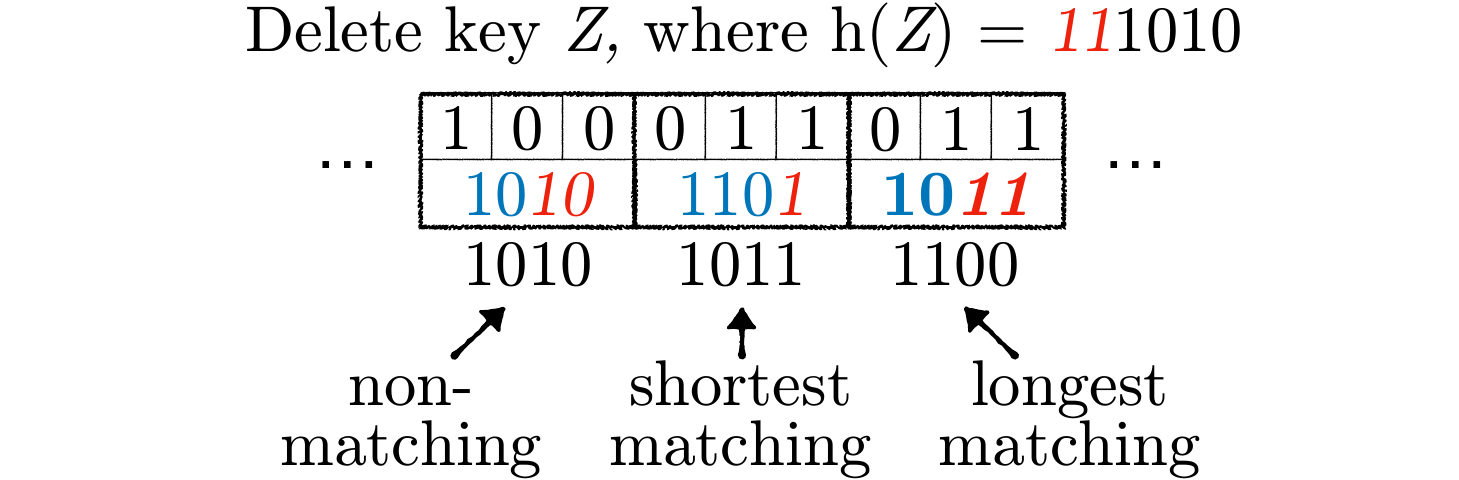}
	\vspace{-2mm}
	\caption{ InfiniFilter deletes the longest matching fingerprints in the target run to prevent future false negatives.   }
	\vspace{-4mm}
	\label{fig:deletes}
\end{figure}

\Paragraph{Deletes} InfiniFilter  deletes a key by removing the longest matching fingerprint in the target run to prevent future false negatives. 
For example, Figure \ref{fig:deletes} illustrates a delete operation to Key~$Z$. The target run for this key begins at Slot 1010 and consists of three slots. While the fingerprint at Slot 1010 does not match, both subsequent fingerprints, which have different lengths, do match. If we remove the shorter matching fingerprint (at Slot~1011) and it happens to belong to a different key $Y$ with a different mother hash from that of Key $Z$ (e.g., $h(Y)=011010$), future false negatives would occur when querying for Key $Y$. In contrast, removing the longest matching fingerprint (at Slot 1100) guarantees that the remaining shorter fingerprint in the run will still match the non-deleted key so that future queries to it do not result in false negatives. 

\Paragraph{Rejuvenation} In many applications (e.g., key-value stores), a positive query to a filter is followed by fetching the corresponding data entry from storage to return it to the user. In case that the key exists (i.e., a true positive), we can rehash it to derive a longer mother hash and thus rejuvenate (i.e., lengthen) the key's fingerprint, in case it was created before the last expansion. Such rejuvenation operations help keep the FPR low. However,  they are only effective when queries target older entries. Similarly to deletes, a rejuvenation operation must lengthen the longest matching fingerprint in a run to prevent  false negatives. 

\Paragraph{Void Entries} After the first $F$ expansions, the oldest fingerprints in InfiniFilter run out of bits. Such entries are referred to as void entries. Both Figures \ref{fig:infinifilter} and \ref{fig:infinifilter_growing} show the creation of a void entry at Slot 101 after the expansion. 
As shown, the unary code fully occupies a slot that contains a void entry. Any query that encounters a void entry in its target run returns a positive. 
Void entries cannot be uniquely remapped to a slot in a larger filter as there is no extra fingerprint bit to transfer to the entry's canonical slot address. 


\renewcommand{\arraystretch}{1.4}
\begin{table*}
	\centering
	\small
		\caption{A comparison of existing filter expansion techniques against Aleph Filter with respect to the data size $N$ and the initial fingerprint length $F$. Aleph Filter provides faster query/delete/rejuvenation operations, and its Predictive Regime requires less memory given an approximation $N_{est}$ of the ultimate data size.  } \label{tab:table1}
	\begin{tabular}{|p{5.0cm}|p{1.5cm}|p{1.0cm}|p{1.7cm}|p{3cm}|p{1.4cm}|}  
		\hline
		& query/ delete/rejuv & insert & false positive rate (FPR) & fingerprint \newline bits / key & max. \newline expansions   \\
		\hline
		Fingerprint Sacrifice  	\cite{Zhang2021, Bender2012}								&  $O(1)$    	   & $O(1)$ 	&   $O(2^{-F} \cdot N)$ &   $F - O( \lg N)$ & $F$ \\
		 InfiniFilter (Fixed-Width Regime)   		 								& $O(\frac{\lg N}{F})$  & $O(1)$ 	&  $  O(2^{-F}   \cdot \lg N )$ &   $F$  & $ 2^{F}$   \\
		 InfiniFilter  (Widening Regime) 				 & $O(\frac{\lg N}{F + \lg \lg N})$  & $O(1)$	&  $O(2^{-F} ) $ &  $F+  O(\lg \lg N )$  & $\infty$   \\
		 		Aleph Filter (Fixed-Width Regime)													&  $O(1)$    	   		& $O(1)$ 	  &   $O( 2^{-F} \cdot \lg N )$ &   $F$ & $ 2^{F}$ \\
		 Aleph Filter (Widening Regime)		   &  $O(1)$    		   & $O(1)$ 	&   $O(2^{-F} )$ &   $F + O(\lg \lg N)$ & $\infty$ \\
		 Aleph Filter (Predictive Regime)		   &  $O(1)$    		   & $O(1)$ 	&   $O(2^{-F} )$ &   $F + O(\lg | \lg (N / N_{est}) |) $ & $\infty$ \\
		\hline
	\end{tabular}
\end{table*}

\begin{figure}[t]
	\includegraphics[width=8.7cm]{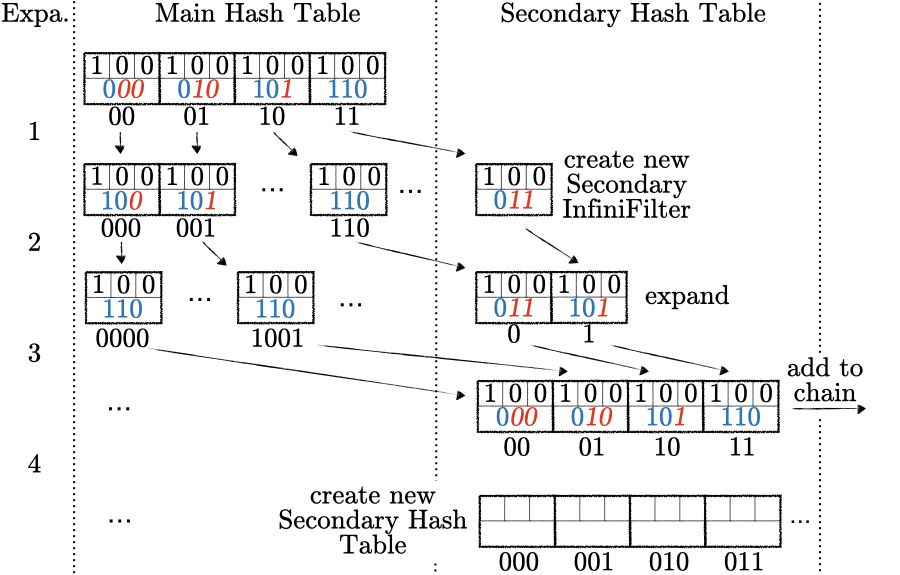}
	\vspace{-2mm}
	\caption{ InfiniFilter transfers void entries into a secondary hash table, which also expands when it reaches capacity. When void entries appear in the secondary hash table, it is appended to a chain of hash tables and a new secondary hash table is allocated.     }
	\vspace{-5mm}
	\label{fig:void}
\end{figure}

\Paragraph{Supporting Infinite Expansions} To support more than $F$ expansions, InfiniFilter  transfers each void entry from the main hash table into a smaller \textit{secondary} hash table, which has an identical structure to that of the main hash table but fewer slots. Figure \ref{fig:void} illustrates this process across four expansions. During the second expansion, for instance, the void entry from canonical Slot 110 of the main hash table is transferred to the secondary hash table. As 110 is also the mother hash of this entry, its least significant bit \mbox{(i.e., 0)} is used as a canonical slot address in the secondary hash table while its more significant two bits (i.e., 11) are used as a fingerprint. 

The secondary hash table  also expands when it reaches capacity. Eventually, the oldest entries in the Secondary hash table become void entries, as shown in Figure \ref{fig:void} by the entry at Slot 11 after the third expansion. As this point, the Secondary InfiniFilter is sealed and appended to a so-called \textit{chain} of auxiliary hash tables, and a new empty secondary hash table is allocated. 
While this design supports an unlimited number of expansions, it slows down  queries, deletes and rejuvenation operations, as they must now traverse potentially all hash tables along the chain. 

%% file: problem.tex
\section{Problem Analysis} \label{sec:problem_analysis}


\Paragraph{Challenge 1: Scaling CPU Costs} The number of hash tables across which InfiniFilter stores its entries determines the CPU costs of queries, deletes and rejuvenation operations. In the Fixed-Width Regime, each hash table stores entries from across $F$ subsequent expansions, and the overall number of expansions is $\log_2(N)$. Hence, there are at most $O(\nicefrac{\lg(N)}{F})$ hash tables. In the Widening Regime, each hash table stores entries from across $F + O(\lg \lg N)$ subsequent expansions on average, and so the number of hash tables is $O( \nicefrac{\lg(N)}{(F + \lg \lg N)}  )$ \cite{Dayan2023}. We summarize these properties in Rows 2 and 3 of Table \ref{tab:table1}. Under both regimes, performance deteriorates as the data grows. Can we reduce the worst-case number of hash tables that queries, deletes, and rejuvenation operations must access? 




\Paragraph{Challenge 2: Alleviating the Memory vs. FPR Contention} 
As shown in Row 2 of Table \ref{tab:table1}, InfiniFilter exhibits a logarithmic FPR if we fix the number of bits per entry.  Alternatively, as shown in Row 3, it exhibits a stable FPR if we increase the number of bits per entry at a doubly logarithmic rate. 
Is it possible to alleviate this scalability contention to achieve an FPR and memory footprint that are both on par with a static filter? 

%% file: main.tex
\section{Aleph Filter} \label{sec:main}

We introduce Aleph Filter, an expandable filter that builds on InfiniFilter to improve its  scalability properties. 
 Section \ref{sec:duplication} shows how Aleph Filter supports queries in $O(1)$ time by keeping and duplicating void entries within the main hash table. Section \ref{sec:analysis} shows analytically that the fraction of duplicated void entries stays small and therefore does not significantly impact the FPR or the maximum number of expansions that the filter supports. 
Sections \ref{sec:deletes} and \ref{sec:rejuv} show how to support deletes and rejuvenation operations in $O(1)$ time {\color{black}without introducing false negatives by lazily identifying and removing void duplicates  corresponding to the entry with the longest matching mother hash.}

Throughout Sections \ref{sec:duplication} to \ref{sec:rejuv}, we analyze Aleph Filter in both the  the Fixed-Width and Widening Regimes and summarize their properties in Rows 4 and 5 of Table \ref{tab:table1}. In Section \ref{sec:predictive}, we  introduce the Predictive Regime. Given a rough, conservative estimate of how much the data will grow, Aleph filter in the Predictive Regime achieves  FPR vs. memory trade-offs that are on par with static filters. We summarize its properties in Row 6 of Table \ref{tab:table1}.

\subsection{Fast Queries by Duplicating Void Entries} \label{sec:duplication}

\begin{figure}[t]
	\includegraphics[width=8.5cm]{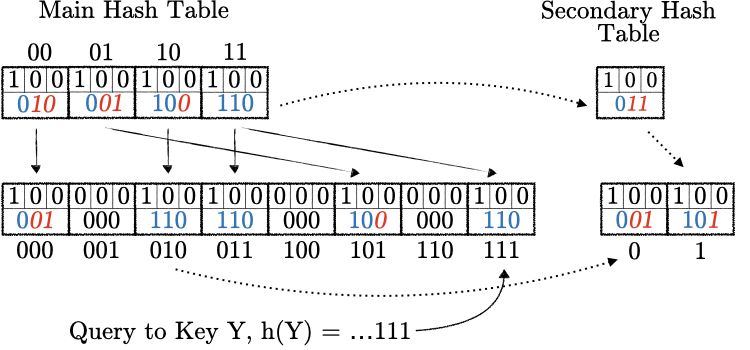}
	\vspace{-6mm}
	\caption{   While expanding, Aleph Filter duplicates each void entry so that one duplicate would still be found in constant time when querying for the original key.   }
	\vspace{-6mm}
	\label{fig:aleph_expansion}
\end{figure}

While expanding, Aleph Filter duplicates each void entry in the main hash table across both canonical slots that it could have mapped to if there were an additional fingerprint bit to sacrifice. 
In Figure~\ref{fig:aleph_expansion}, for example, there is a void entry at Slot 11 of the main hash table before the expansion.  During the expansion, Aleph Filter duplicates it across Canonical Slots 011 and 111.  In the next expansion, each of these duplicates will be duplicated again, resulting in four duplicates at Canonical Slots 0011, 0111, 1011, and 1111. 

As a result of duplicating void entries, a query targeting any old entry returns a positive after one access to the main hash table. Figure \ref{fig:aleph_expansion} illustrates a query to Key $Y$, which corresponds to the void entry that was duplicated. The query visits canonical slot 111 (based on the first three bits of the key's mother hash), finds a void entry and returns a positive.  Had Key Y's mother hash started with 011, the query would have found the other void duplicate at Slot 011 and also terminated after one hash table access. {\color{black} As there is a void entry in every possible run that would have contained the entry if we had all bits of its mother hash, a query to the the original key always returns a positive. Hence, no false negatives can occur. }


Another result of duplicating void entries is that any query to a non-existing key terminates after one access to the main hash table. The reason is that the main hash table contains either a fingerprint or a void entry in any existing key's canonical slot. This means that if we do not find a matching entry in $O(1)$ time, the target entry is guaranteed not to exist. 

Hence, all queries to Aleph Filter are processed in worst-case $O(1)$ time. This is an improvement over InfiniFilter, where queries targeting non-existing or older keys must search an increasing number of hash tables as the data grows. 

\subsection{Analysis} \label{sec:analysis}

Duplicating void entries in the main hash brings up two plausible concerns. 
The first is whether duplicated void entries significantly increase the FPR, seeing as a query that encounters a void entry immediately returns a positive.  
The second is whether duplicated void entries take up significant space within the main hash table. 
This section shows that the proportion of void entries stays small and therefore does not significantly increase the FPR or cause the hash table to fill up prematurely. 

\Paragraph{Generational Distribution} Aleph Filter doubles in capacity during each expansion. This means that keys inserted zero, one or two expansions ago comprise approximately a half, a quarter, or an eighth of the data set, respectively, and so on. More generally, consider the Set $s_j$ of keys inserted in-between the  $j^{\text{th}}$ and $(j+1)^{\text{th}}$ expansions, i.e., in Generation $j$. Suppose we are now in Generation $X$ ($X \ge j$), i.e., before Expansion $X+1$. 
Equation \ref{eq:freq} approximates the  size of Set $s_j$ as a fraction of the whole key set. 
\begin{equation} \label{eq:freq} \small
	f(j) \approx 2^{-X +j -1}
\end{equation}

\Paragraph{FPR in the Fixed-Width Regime} 
Consider some Generation $j$ of non-void entries (i.e., $X - j < F$). These entries' fingerprints each consist of $F - (X - j)$ bits, since they are initialized with $F$ bits and lose one bit in each expansion. 
The probability for a query to encounter such an entry in a given slot and falsely match its fingerprint  is $\alpha \cdot f(j) \cdot 2^{-F + X - j}  = \alpha \cdot  2^{-F-1}$. 

Now consider a Generation $j$ of void entries (i.e., \mbox{$X - j \ge F$}). There are $2^{-F + X - j}$ duplicates per entry since the number of duplicates increases by a factor of two in each expansion. 
The fraction of slots occupied by void duplicates originating from Generation $j$ is $\alpha \cdot f(j) \cdot 2^{-F + X - j} =  \alpha \cdot 2^{-F-1}$. Hence, the probability of encountering a void duplicate  from Generation $j$ during a query and thus returning a false positive is $\alpha \cdot 2^{-F-1}$. 

As shown, the  contribution of each generation of entries to the FPR is equal (i.e., $\alpha \cdot 2^{-F-1}$). 
For non-void entries, this is because each time their fingerprints lose one bit, their fraction in the overall filter halves. For void entries, this is because they always return a positive, and their fraction of void entries emanating from a given generation stays fixed since they duplicate. 
Equation \ref{eq:FPR_fixed_width2}  expresses the overall FPR after $X$ expansions. We see that Aleph Filter has approximately the same FPR as InfiniFilter in the Fixed-Width Regime. 
\begin{equation} \label{eq:FPR_fixed_width2} \small
	\small
	\begin{split}
	FPR &\approx \sum_{j=0}^{X + 1} \alpha \cdot  2^{-F-1}   \lesssim \alpha \cdot (log_2(N)+2) \cdot 2^{-F-1} \\	
	\end{split}
\end{equation}


\Paragraph{FPR in the Widening Regime} In the Widening Regime, entries in Generation $j$ are assigned fingerprints of $\ell(j) = F + 2 \cdot \log_2(j+1)$ bits, and they lose one bit in each expansion. 

Let us consider a generation Generation~$j$ of non-void entries (i.e., $X - j < \ell(j)$). By Generation $X$, the fingerprints of Generation~$j$ will have shrunk by $X-j$ bits to $\ell(j) - (X - j)$ bits. 
The probability for a query to encounter an entry from Generation~$j$ in a given slot and falsely match its fingerprint is therefore  {$\alpha \cdot f(j) \cdot 2^{- \ell(j) + X - j }$}. 

Let us now suppose Generation $j$ consists of void entries (i.e., \mbox{$X - j \ge \ell(j)$}). The number of void duplicates for each entry is $2^{- \ell(j) + X - j }$. The probability of encountering a void duplicate originating from this generation is therefore the same expression: {$\alpha \cdot f(j) \cdot 2^{- \ell(j) + X - j }$. 

Equation \ref{eq:memory5} sums up this expression across all generations to derive the FPR. The derivation uses a well-known identity that the sum of the reciprocals of the square numbers (i.e., $\Sigma_{i=0}^{\infty} i^{-2}$) converges to $\nicefrac{\pi^2}{6}$. 
The result is that Aleph Filter's FPR matches InfiniFilter's FPR in the Widening Regime. 
\begin{equation} \label{eq:memory5} \small
	\small
	\begin{split}
	FPR & \approx \sum_{j=0}^{X+1} \alpha \cdot  f(j) \cdot 2^{ - \ell(j) + X - j  }  \\[-2mm]
	& \approx \alpha \cdot 2^{-F-1} \cdot \sum_{j=0}^{i}  \cdot \frac{1}{ (j+1)^2} \lesssim \alpha \cdot 2^{-F-1} \cdot \frac{\pi^2}{6} \lesssim \alpha \cdot 2^{-F}
	\end{split}
\end{equation}

\Paragraph{Expansion Limit in Fixed-Width Regime} 
We saw above that the fraction of slots in the filter that are occupied by void duplicates of entries created in generation $j$ is $\alpha \cdot f(j) \cdot 2^{-F + X - j} =  \alpha \cdot 2^{-F-1}$. The number of generations of void entries is $X-F + 1$. Hence, the fraction of slots occupied by void duplicates by the time we reach Generation $X$ ($X \ge F$) is $\gamma(X) = \alpha \cdot 2^{-F-1} \cdot (X-F + 1)$. 
We may now ask at which point the number of duplicates takes up half the space in the filter, meaning that even after we expand, void duplicates fully take up the expanded capacity so there is no space for new insertions. We obtain it by setting $\alpha$ to one, equating $\gamma(X)$ to \nicefrac{1}{2}, and solving for $X$ to obtain $F + 2^F - 1$. Recall from Section \ref{sec:background_sec} that the Fixed-Width Regime only supports $2^{F}$ expansions anyways (as at this point the FPR reaches one and the filter becomes useless). Hence, Aleph Filter does not reduce the maximum number of supported expansions in the Fixed-Width Regime.

\Paragraph{Expansion Limit in Widening Regime} 
As we saw, the fraction of slots  occupied by void duplicates from Generation~$j$ is  $\alpha \cdot f(j) \cdot 2^{- \ell(j) + X - j }$. 
Let $v$ denote the total number of generations of void entries. 
The overall fraction of slots occupied by void entries is  $\sum_{j=0}^{v} \alpha \cdot f(j) \cdot 2^{- \ell(j) + X - j }$. 
This expression is subsumed by Equation \ref{eq:memory5} and is therefore lower than $\alpha \cdot 2^{-F}$. 
This establishes that the fraction of slots occupied by void entries converges to a small constant as the filter expands. 
The intuition is that newer entries are assigned longer fingerprints, and so it takes them increasingly longer to become void and start duplicating. 
Hence, Aleph Filter supports an infinite number of expansions in the Widening Regime.

\subsection{Fast Deletes Using Tombstones} \label{sec:deletes}

Duplicating void entries makes delete operations  intricate to handle. The reason is that every duplicate of a void entry that we wish to delete must be identified and removed.  The challenge is doing so in constant time, without using a significant amount of additional metadata, {\color{black} and without creating the possibility of false negatives.}


In the simple case that a delete operation is targeting a run with at least one matching non-void entry, the entry with the longest matching fingerprint will be removed as shown in Section~\ref{sec:background_sec}. This section focuses on processing deletes when the only matching entries in the target run are void entries.

\begin{figure}[t]
	\includegraphics[width=8cm]{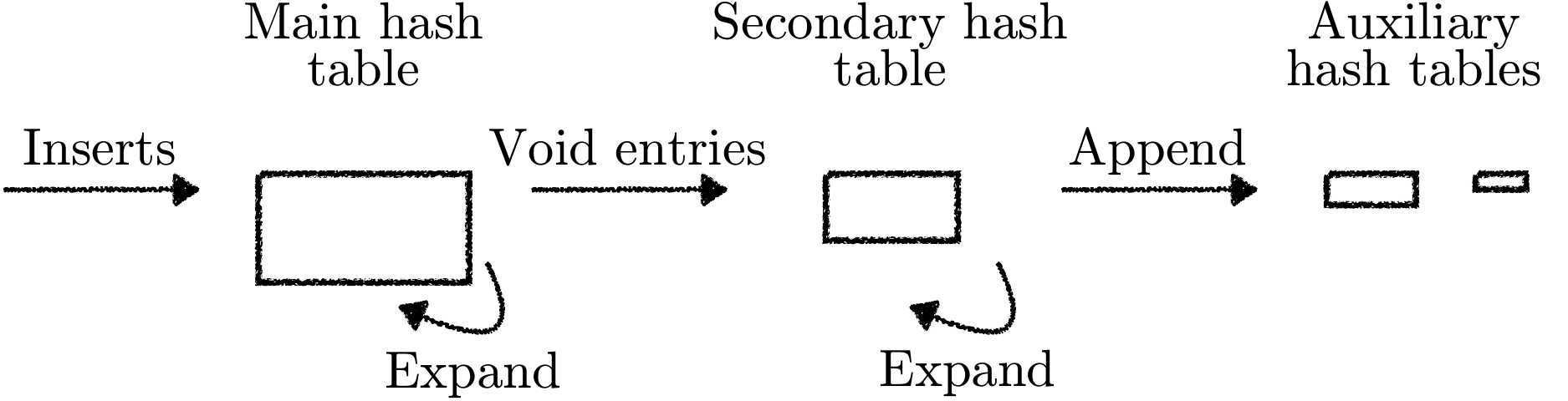}
	\vspace{-2mm}
	\caption{ Aleph filter stores the mother hashes of all void entries along the secondary and auxiliary hash tables.      }
	\vspace{-5mm}
	\label{fig:chaining}
\end{figure}


\Paragraph{Identifying a Void Entry's Duplicates} An entry becomes void when the length of the mother hash that it was assigned when it was inserted matches the logarithm base two of the number of slots in the filter. In every subsequent expansion, the number of duplicates for the entry and the number of slots in the filter both multiply by a factor of two. If the entry's original mother hash consists of $b$ bits and the filter currently comprises~$2^k$ slots, then the entry participated in $k-b$ expansions since it became a void entry. Therefore, it has $2^{k-b}$ void duplicates. 
For example, suppose the filter consists of $2^6$ slots and the mother hash we wish to remove is 0011. This mother hash consists of $b=4$ bits while the power of the size of the filter is $k=6$, meaning that its number of void duplicates is $2^{6-4} = 4$. 

We can also infer which canonical slots contain these void duplicates by using the mother hash of the original entry as the least significant $b$ bits of their addresses, and applying all possible permutations to the remaining $k-b$ bits. 
In the example just given, we use the mother hash 0011 as the least significant bits and permute the remaining two bits to obtain the following four slots addresses: \textbf{00}0011, \textbf{01}0011, \textbf{10}0011, \textbf{11}0011.  To execute the delete correctly, we would need to remove a void duplicate from each one of these canonical slots. 
The question becomes how to efficiently store and retrieve the mother hash of any void entry so that we can identify and remove all of its void duplicates? 

\Paragraph{Secondary Hash Table} Similarly to the Chained InfiniFilter, Aleph filter adds the mother hash of any entry that turns void into a secondary hash table, which has an identical structure to that of the main hash table yet fewer slots. 
Figure \ref{fig:aleph_expansion} illustrates an example. Before the expansion, the secondary hash table stores only the mother hash of the void entry at Slot 11 of the main hash table. The secondary hash table then expands alongside the main hash table. After the expansion, another entry turns void at Slot 010 of the main hash table, and so its mother hash is stored in the Secondary hash table.  Note that while a void entry may have multiple duplicates in the main hash table (e.g., at Slots 011 and 111), its mother hash is only stored once in the secondary hash table.




\Paragraph{Auxiliary Hash Tables} As the Secondary Hash Table fills up, it must expand as well. Expanding it entails transferring one bit from each entry's fingerprint to its canonical slot address in the expanded hash table. 
In Figure \ref{fig:aleph_expansion}, for example, the entry at Slot~1 of the secondary hash table after the expansion corresponds to the entry with fingerprint 11 before the expansion, since the least significant bit of its fingerprint is repurposed as the most significant bit of its slot address. Eventually, void entries appear in the secondary hash table as well. At this point, we seal the secondary hash table and add it to a chain of auxiliary hash tables. A new empty secondary hash table is  then allocated. Figure~\ref{fig:chaining} illustrates the high-level workflow. 
This architecture is similar to that of the Chained InfiniFilter from Section~\ref{sec:background_sec}. The core difference is that InfiniFilter traverses the secondary and auxiliary hash tables to process queries and deletes while Aleph Filter does not. 

\Paragraph{Tombstones} A delete operation commences by modifying a void entry in the canonical slot of the key to be deleted into a tombstone at the main hash table. This causes subsequent queries to the deleted key to likely return a negative (unless there is some other void entry in the slot, which would lead to a false positive). To encode a tombstone, we employ a special bit string of all 1s. Figure \ref{fig:tombstone} shows an example of a delete operation of Key $X$ landing at Canonical Slot 101, where the only matching entry is a void entry. The first step is changing the content of the slot from a void entry encoding (i.e., 1110) into a tombstone (i.e., 1111). As a result, subsequent queries to Key $X$ in the example will now return a negative.

\begin{figure}[t]
	\includegraphics[width=8.4cm]{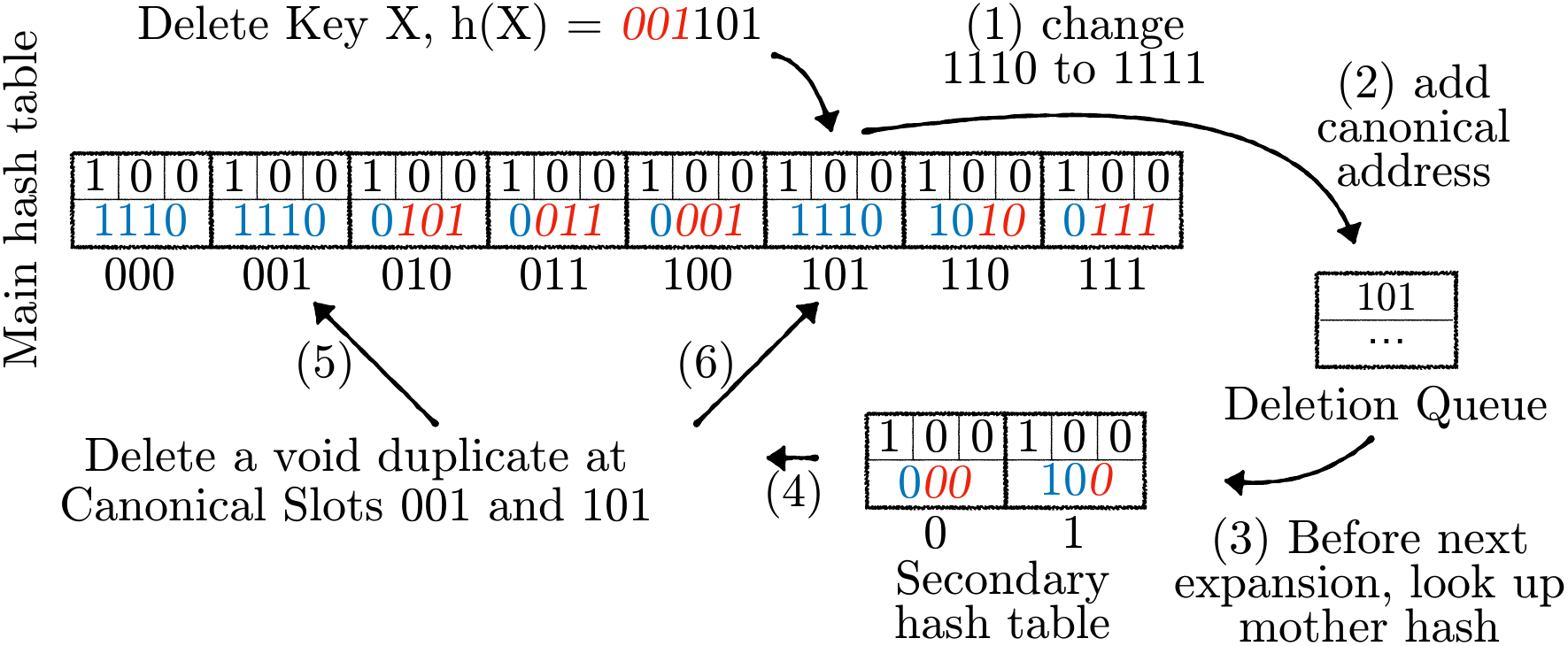}
	\vspace{-3mm}
	\caption{  Aleph filter deletes a void entry by first transforming it into a tombstone and storing the canonical address in a deletion queue. Before the next expansion, void duplicates are identified and removed.    }
	\vspace{-4mm}
	\label{fig:tombstone}
\end{figure}

\Paragraph{Deferred Removal of Duplicates} 
The second step of a delete operation is to add the canonical slot of a void entry to be removed into a deletion queue, which is structured as an append-only array. 
Right before the next expansion, Aleph Filter pops one canonical slot address at a time from the deletion queue. It uses this address as a search key to traverse the secondary and auxiliary hash tables from largest to smallest. The search terminates as soon as it finds a matching entry. Since larger hash tables store longer mother hashes and we search the hash tables from largest to smallest, the first matching entry that we find corresponds to the longest matching mother hash. Based on this mother hash, we identify all duplicates and remove them from the main hash table\footnote{Note that if many such deletes of void entries take place before the next expansion and cause the filter utilization to drop significantly below the expansion threshold again, we delay the expansion until utilization reaches the threshold again. }.



In Figure \ref{fig:tombstone}, for example, we pop Slot 101 from the deletion queue and use it as a search key to probe the secondary hash table. We find a matching entry at Slot 1 with fingerprint of 0. We concatenate these to obtain a longest matching mother hash of 01. Since the main hash table currently has $2^3$ slots while the longest matching mother hash consists of two bits, we infer that the  void entry has $2^{3-2}=2$ duplicates in the main hash table, and that their canonical slots are \textbf{0}01 and \textbf{1}01. We proceed to delete one void entry from each of these canonical slots. Finally, we also delete the longest matching mother hash from the secondary hash table. 






\Paragraph{Preventing False Negatives} Figure \ref{fig:aleph_deletes2} shows a more complex example of a delete operation where there are two void entries in the target canonical slot. When we visit the Secondary Hash Table, we find a run containing two mother hashes of different lengths for these void entries: 00 and 000. The different lengths of these mother hashes indicate that one of the void entries has two duplicates in the filter while the other has one. There is now a question of whether to remove the entry with the single void entry or the one with the two void duplicates.

Let us first suppose the entry with the smaller mother hash (and hence more duplicates) is removed. If this entry happens to correspond to a different key Y with a different extended mother hash than that of key X (e.g., $h(Y)=...100$),  we would get false negatives later when querying for Key $Y$. Particularly, a query for Key $Y$ would reach Slot 100, fail to find a matching entry, and return a false negative. To prevent false negatives, we must delete the void entry with the fewest duplicates. As before, this entry corresponds to the one with the longest matching mother hash. This ensures that the remaining void entry's duplicates will still match whichever entry still exists.  
Hence, in Figure \ref{fig:aleph_deletes2}, one void entry is removed from Slot 001 of the main hash table, and the entry with Slot 0 is removed from the secondary hash table. 

\begin{figure}[t]
	\includegraphics[width=8.3cm]{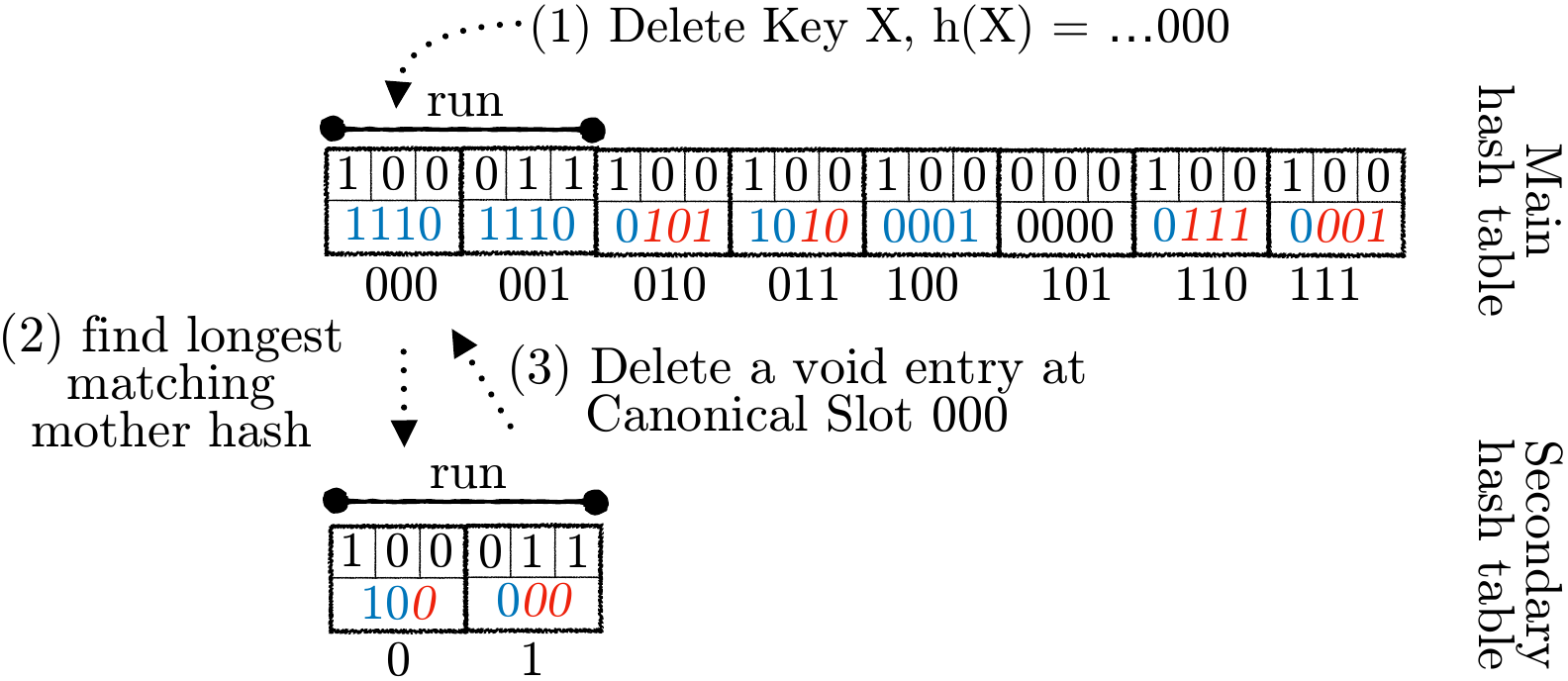}
	\vspace{-3mm}
	\caption{ Aleph filter identifies which void duplicates to remove based on the longest matching mother hash of the deleted key.   }
	\vspace{-4mm}
	\label{fig:aleph_deletes2}
\end{figure}


\Paragraph{Computational Analysis} Turning a void entry into a tombstone and adding its address to the deletion queue take $O(1)$ time to execute.
The cost of identifying and removing all duplicates for a given void entry is deferred and incurred right before the next expansion. To quantify this cost in the worst-case, suppose the user deletes all void entries at once. The dominating part of the cost is to remove each individual duplicate from the main filter, while the cost of retrieving every mother hash from the secondary and auxiliary hash tables is a lower-order term (i.e., $O(\nicefrac{X}{F})$).  In Section~\ref{sec:analysis}, we saw that there are at most $O( 2^{-F} \cdot X \cdot N )$ void duplicates in the main hash table in the Fixed-Width Regime. Each of them takes constant time to remove.  Hence, the overall amount of work is $O( 2^{-F} \cdot X \cdot N )$. 
As this work is performed after $N$ insertions, it remains sub-constant as long as $X < 2^{F}$. Since this is the maximum number of supported expansions anyways, the cost of removing void duplicates is amortized constant for the filter's whole lifetime. 





In the Widening Regime, we saw in Section \ref{sec:analysis} that there are at most $O( 2^{-F}  \cdot N )$ void duplicates in the main hash table. Removing each of them after $N$ insertions entails $O(2^{-F})$ additional overhead per insertion. Hence, Aleph Filter supports deletes in constant time while being able to expand indefinitely in the Widening Regime.

\Paragraph{Memory Analysis} The secondary and auxiliary hash tables store at most $N \cdot 2^{-F}$ mother hashes altogether. The size of these hash tables is therefore smaller by a factor of at least $2^{-F}$ from the main hash table and therefore does not take up much memory. 




\subsection{Fast Rejuvenation Operations} \label{sec:rejuv}

In Section \ref{sec:background_sec}, we saw that a rejuvenation operation rehashes a queried key after retrieving it from storage to lengthen the longest matching fingerprint in the target run. This helps reduce the FPR. 
In Aleph Filter, a rejuvenation operation can be trickier to handle if the only matching entry in the target run is a void entry. In this case, we must also eliminate its void duplicates. 

Aleph Filter handles the case where the target run only contains matching void entries by immediately rejuvenating one void entry into the full fingerprint of the queried key. It also adds the corresponding canonical slot into a Rejuvenation Queue (similar to the Deletion Queue from Section~\ref{sec:deletes}). Just before the next expansion, Aleph filter pops one address at a time from the Rejuvenation Queue. For each address, it finds the longest matching mother hash in the Secondary or Auxiliary Hash Tables. It compares the length of this mother hash to the log of the number of slots in the main hash table to infer how many void duplicates correspond to this mother hash and what their locations are. It then removes each of these void duplicates. This process is identical to how deletes are processed with the only exception that a void duplicate need not be removed from the queried key's canonical slot, as the void entry there has already been transformed into a full fingerprint upfront. 

Figure \ref{fig:rejuvenation} shows an example of rejuvenating a void entry at Canonical Slot 100. First, Aleph Filter transforms the void entry into a full fingerprint and adds the canonical slot address to the Rejuvenation Queue. 
Before the next expansion, Aleph filter pops address 100 from the Rejuvenation Queue and uses it as a search key to query the Secondary Hash Table. It finds the longest matching fingerprint of 0 at Slot 0 (the other mother hash in this run is 000, which doesn't match 100 along the third bit). Hence, the longest matching mother hash is 00.
This implies that the void entry had been duplicated twice at Canonical Slots \textbf{0}00 and \textbf{1}00 of the main hash table. However, since this is a rejuvenation operation, we know we have already replaced one void entry by a fingerprint at Canonical Slot 100, so the only remaining duplicate to remove is at Canonical Slot 000. Finally, we also remove the longest matching mother hash from the Secondary Hash Table. 

Rejuvenation operations do a constant amount of work upfront while deferring and amortizing the removal of duplicates to the next expansion. Hence, their cost is $O(1)$.

\begin{figure}[t]
	\includegraphics[width=8.3cm]{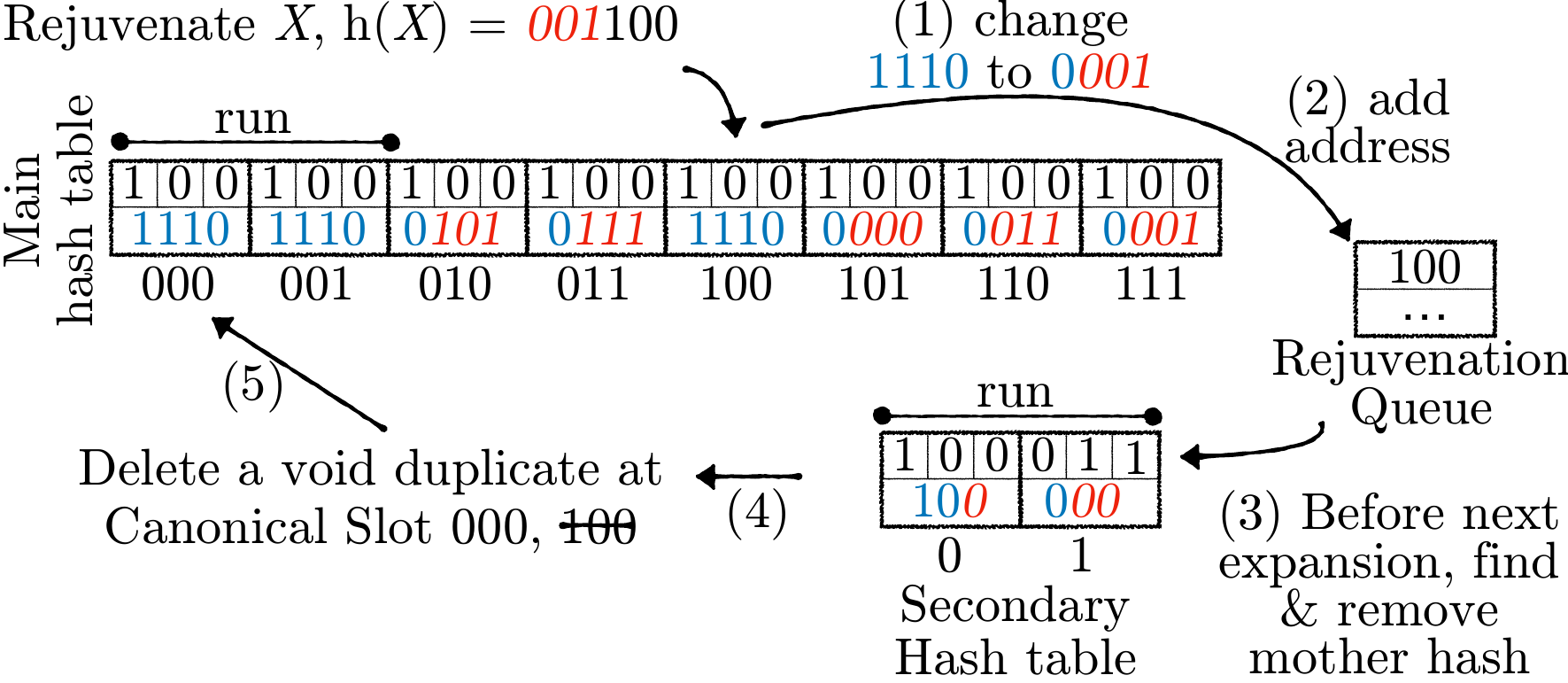}
	\vspace{-3mm}
	\caption{  Aleph filter rejuvenates a void entry by turning it into a full fingerprint and adding the canonical slot to a rejuvenation queue. It removes duplicates lazily.    }
	\vspace{-4mm}
	\label{fig:rejuvenation}
\end{figure}

\section{Predictive Regime} \label{sec:predictive}

So far, we have seen two complementary methods of scaling the FPR as the data grows. Rejuvenation operations lengthen fingerprints to reduce the FPR, but they are only effective when the query workload is targeting older entries (i.e., with shorter fingerprints). On the other hand, the Widening Regime scales the FPR by increasing the fingerprint length assigned to newer entries, yet the cost is a higher memory footprint of $F + O(\lg \lg N)$ bits / entry. 
This begs the question of whether there are  ways to fix the FPR without relying on rejuvenation operations or using more memory.  
This section introduces the Predictive Regime to address this challenge.  The Predictive Regime is an orthogonal contribution to Aleph Filter. It is also applicable to other expandable filters (e.g., InfiniFilter \cite{Dayan2023}). 

The Predictive Regime takes as a parameter an estimate  from the user of how much the data  will grow. We denote this estimate as $N_{est}$, and it is measured as the ratio between what we think the final data size will be to the initial filter capacity. 
Using this estimate, the Predictive Regime uses Equation \ref{eq:predictive_FP2}  to assign fingerprints of length $\ell(j)$ to entries inserted at Generation $j$. The term $X_{est} = \log_2( N_{est} )$ refers to the number of expansions before reaching the estimate. 

At Generation 0, Equation \ref{eq:predictive_FP2} assigns longer fingerprints of length $F + \lceil 2 \cdot \log(X_{est} - 1) \rceil$ bits. 
As the data size grows towards the estimate ($1 \leq  j 	\leq X_{est}$), it assigns an equal or shorter fingerprint length to every subsequent generation. 
When we reach the data size estimate (i.e., $j=X_{est}+1$), the Equation assigns  fingerprints of length $F$ bits, while all fingerprints assigned in previous generations will have shrunk to at most $F$ bits. 
Note that this is in contrast to the Widening Regime, in which we begin with fingerprints of length~$F$ bits and assign monotonically longer fingerprints as the filter expands. 
The max function in Equation \ref{eq:predictive_FP2} keeps the inner part of the logarithm non-zero so that the equation is defined for all $j \ge 0$. 
\begin{equation} \label{eq:predictive_FP2} \small
	\ell(j) = F + 2 \cdot \lceil  \log_2(  \max( | X_{est} - 1 - j | , 1) )   \rceil 
\end{equation}

After surpassing the data size estimate (i.e., $j> X_{est}+1$), the Predictive Regime assumes the Widening Regime's behavior. The absolute value function within the logarithm in Equation \ref{eq:predictive_FP2} sets  increasing fingerprint lengths to subsequent generations to keep the FPR stable. Note that when we set $N_{est} = 1$ (implying $X_{est} = 0$), the Predictive Regime is identical to the Widening Regime from the get-go. Generally, the memory complexity for the Predictive Regime is $F + O(\lg | \lg (N / N_{est}) |) $.

 \Paragraph{Visualization} Figure \ref{fig:predictive} illustrates how the fingerprint length assigned to newer entries first drops during the first $X_{est}$ expansions and then increases again afterwards. The Predictive Regime initially  requires more memory than the Widening Regime. As the data grows, however, the Predictive Regime comes to improve upon the Widening Regime across the board, especially when the real data size is close to the estimate. This is a good trade-off; it is better to take up fewer bits per entry when the data is large rather than when it is small. 
 
 \Paragraph{FPR Analysis} The analysis of the FPR with the Predictive Regime is similar to the analysis of the Widening Regime from Section~\ref{sec:analysis}. The contribution of any generation $0 \le j \le X$ to the FPR is $\alpha \cdot f(j) \cdot 2^{\ell(j) - (X-j)} $. Equation \ref{eq:predictive_FPR} sums this up across all generations to obtain the overall FPR. 
\begin{equation} \label{eq:predictive_FPR} \small
	\small
		FPR = \sum_{j=0}^{X+1} \alpha \cdot f(j) \cdot 2^{\ell(j) - (X-j)}
\end{equation}

By plugging in $X_{est}$ for $X$ in Equation  \ref{eq:predictive_FPR} and simplifying, we obtain the maximum FPR until the moment we reach the data size estimate. This turns out to be at most $2^{-F}$ by the same analysis we saw in Section \ref{sec:analysis} for the FPR in the Widening Regime. At this point, all fingerprints consist of at most $F$ bits. Hence, the memory vs.  FPR trade-off is on par with a static filter by the time we reach the data size estimate. 

By plugging infinity for $X$ in Equation  \ref{eq:predictive_FPR} and simplifying, we obtain a maximum FPR of $2^{-F+1}$ as we surpass the data size estimate.  The intuition is that the left-hand and right-hand curves of the Predictive Regime in Figure \ref{fig:predictive} each contribute an additive factor of $2^{-F}$. 
It is possible to use one extra  bit in advance to maintain a given FPR target as we surpass the data size estimate. 

\begin{figure}[t]
	\includegraphics[width=8cm]{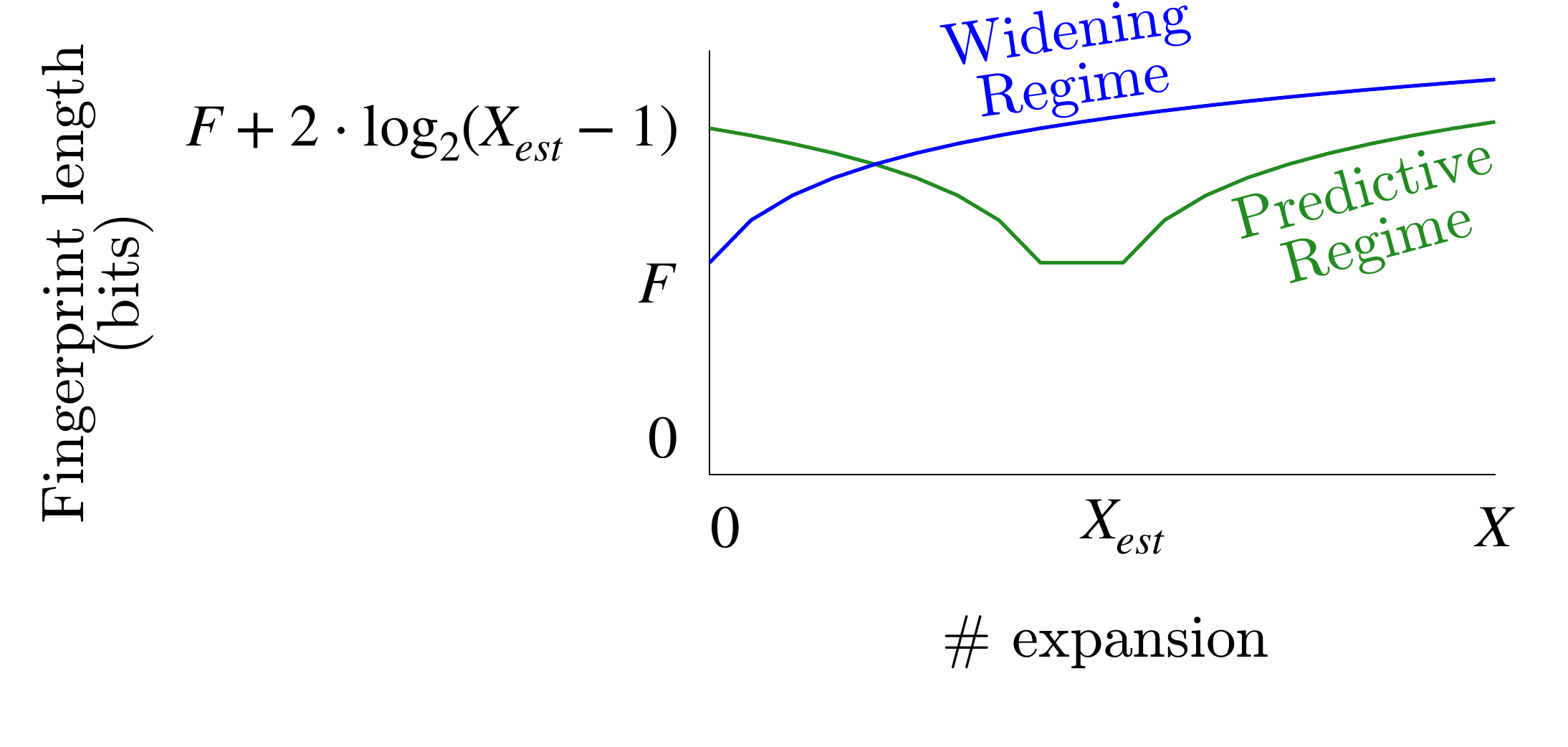}
	\vspace{-6mm}
	\caption{  Given a rough estimate of how much the data size will grow, Aleph filter requires far fewer bits per entry by the time we reach the estimate and pass it.   }
	\vspace{-4mm}
	\label{fig:predictive}
\end{figure}

%% file: evaluation.tex
 \begin{figure*}[!t]
	\includegraphics[scale=0.4]{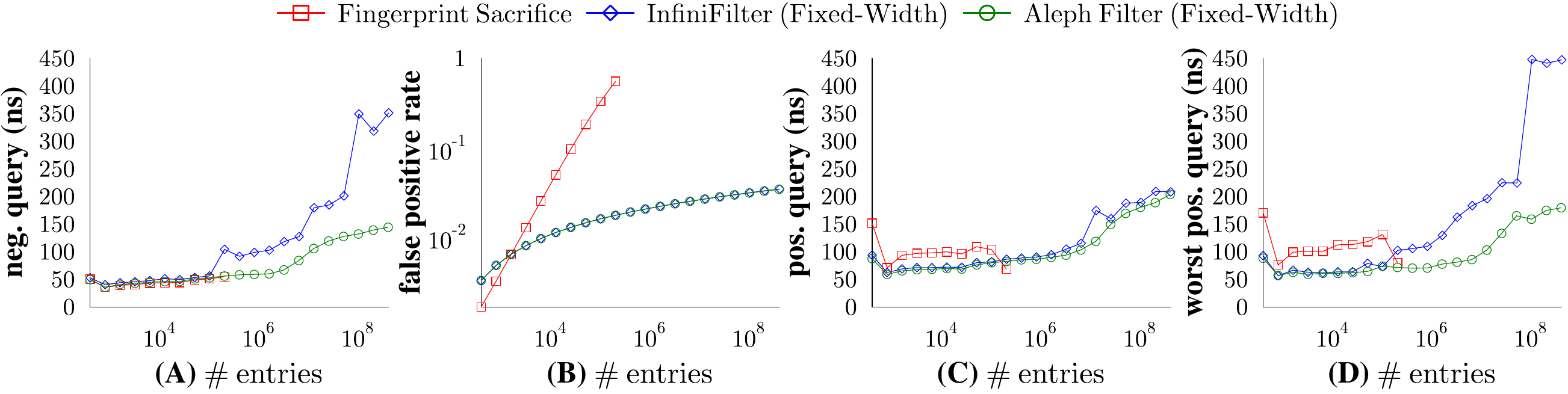}
	\vspace{-3mm}
	\caption{ Aleph Filter exhibits faster worst-case queries while matching InfiniFilter in terms of the false positive rate.  }
	\vspace{-3mm}
	\label{fig:exp1}
\end{figure*}

\begin{figure*}[!t]
	\includegraphics[scale=0.4]{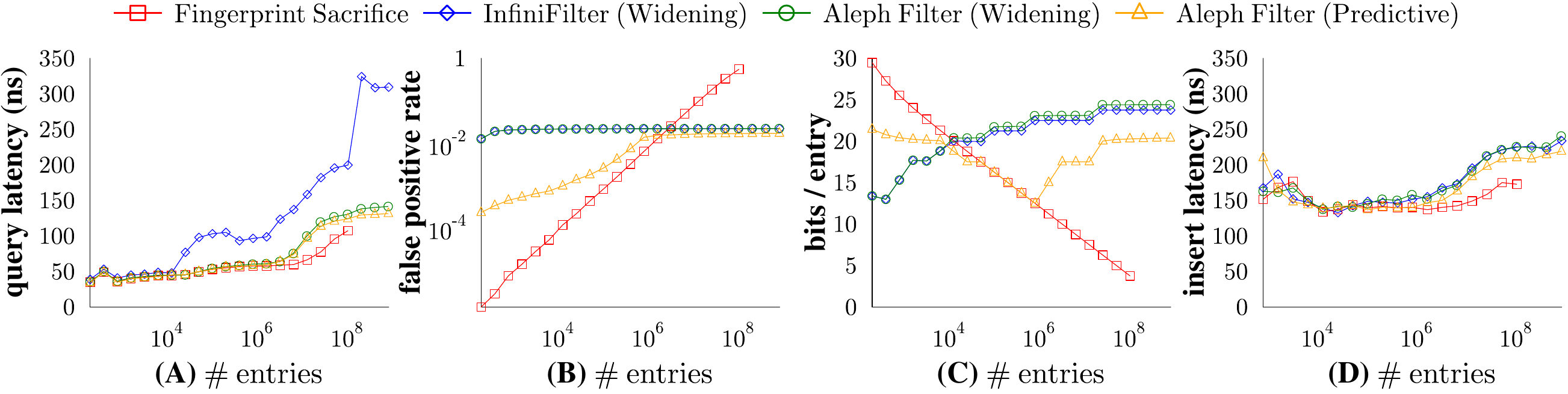}
	\vspace{-4mm}
	\caption{  The Predictive Regime is best when we have a lower bound on how much the data will grow. 	}
	\vspace{-1mm}
	\label{fig:pred}
\end{figure*}

\section{Evaluation}  \label{sec:eval}

We evaluate Aleph Filter against the Fingerprint Sacrifice (FS) method and InfiniFilter, the state-of-the-art techniques for expanding filters, which were described and analyzed in Sections~\ref{sec:background_sec} and ~\ref{sec:problem_analysis}. 

\Paragraph{Implementation} 
We built Aleph filter  as a fork and subclass of InfiniFilter to reuse its core machinery   (e.g., for parsing slots, migrating entries, etc.). All baselines inherit from the same Quotient Filter base class. Reusing code across the baselines means that any performance differences  arise due to their expansion algorithms rather than implementation idiosyncrasies. We employ version 11.0.16 of the Java compiler. 

\begin{figure}[!t]
	\includegraphics[scale=0.36]{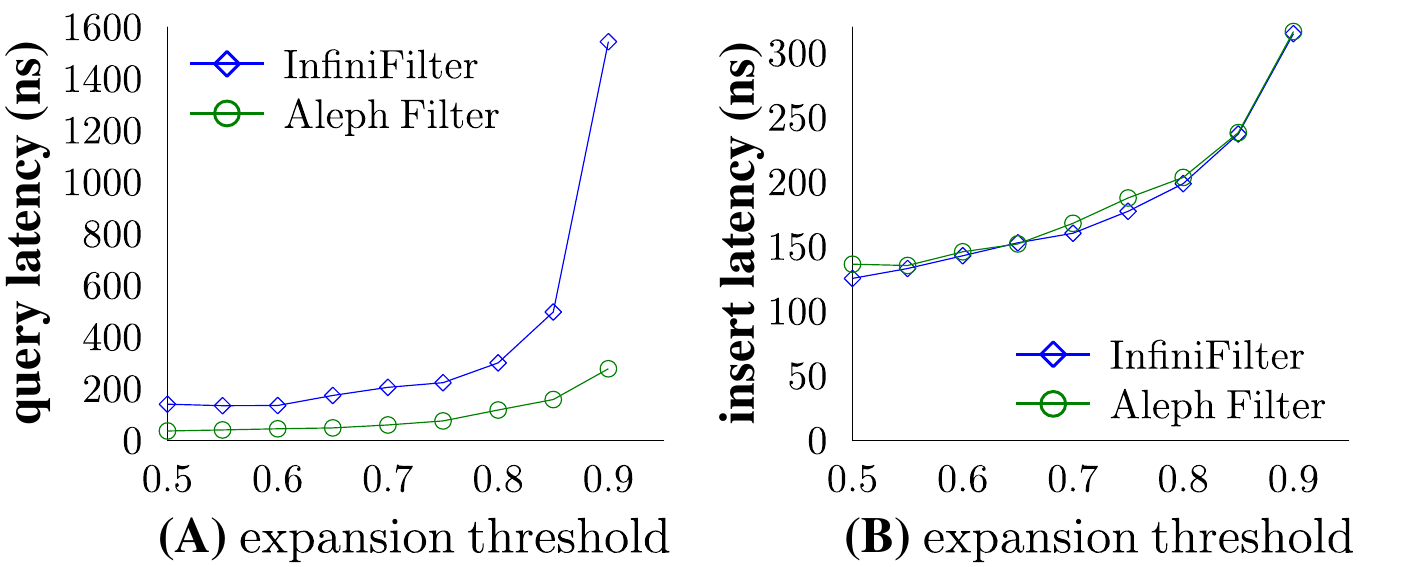}
	\vspace{-4mm}
	\caption{  \color{black} The expansion threshold allows trading between space and performance. Aleph Filter exhibits more robust query performance as we vary this parameter.     }
	\vspace{-1mm}
	\label{fig:threshold}
\end{figure}

\begin{figure}[!t]
	\includegraphics[scale=0.36]{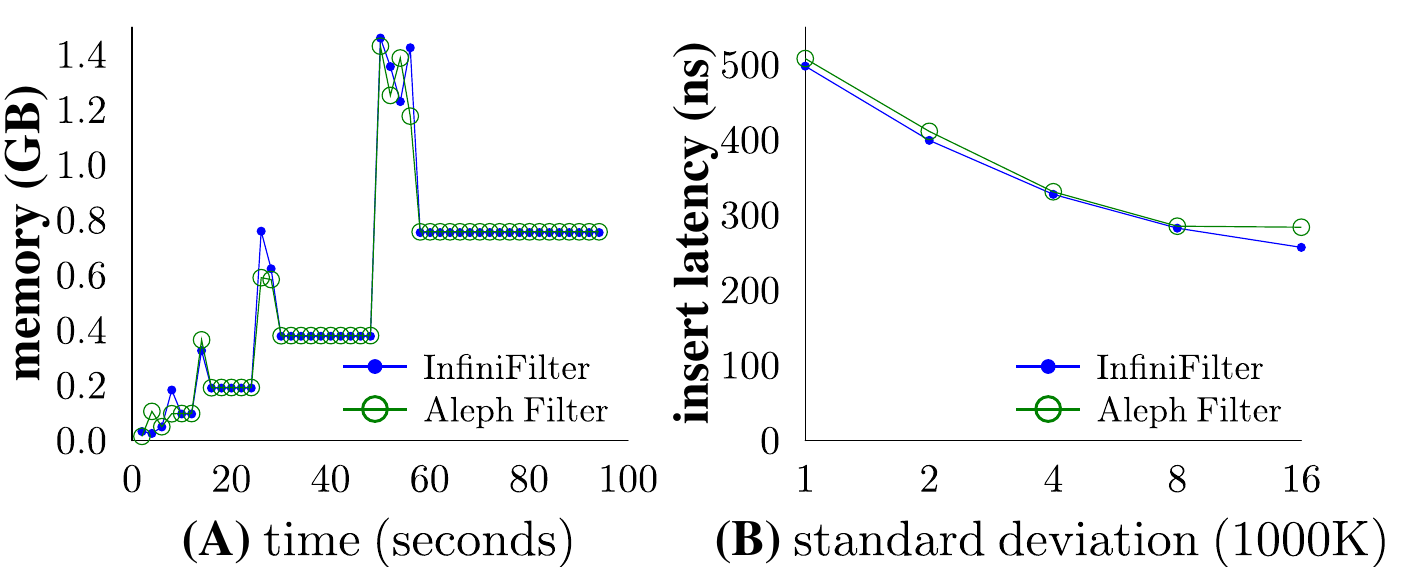}
	\vspace{-3mm}
	\caption{   \color{black} All baselines exhibit space spikes while expanding (Part A) and slowdown due to skewed insertions (Part B).    }
	\vspace{-3mm}
	\label{fig:memory_skew}
\end{figure}

{
\color{black}
\Paragraph{Workload Description} All experimental trials begin with a filter consisting of $256$ canonical slots and issuing insertions, causing the filter to expand multiple times. Unless otherwise mentioned, all queries are issued to a given baseline right before the next expansion to measure the worst-case query performance (when clusters are longest). Other than Figures \ref{fig:exp1} Parts (C) and (D), the queries in all experiments target non-existing keys. Each query experiment issues 10k queries and averages their latency. Aside from Figure~\ref{fig:memory_skew} Part (B), all experiments employ uniformly random workloads. For all baselines, we use java.util.Random to generate random keys and xxhash~\cite{XXHash} as the hash function. All data keys are eight-byte integers before being hashed. Each baseline expands when 80\% of the hash table is occupied, though we vary this threshold in Figure~\ref{fig:threshold}. Figure \ref{fig:pred} focuses on the Widening and Predictive Regimes, while all other experiments use the Fixed-Width Regime with 12-bit slots. 
}

\Paragraph{Hardware} Our system is equipped with two Intel Xeon E5-2690v4 processors, each running at 2.6 GHz with 14 cores and two hyper-threads per core. The machine contains 512GB of RAM, 35MB of L3 cache, 256KB of L2 cache, and 32KB of L1 cache. Storage includes two 960GB SSDs and four 1.8TB HDDs, though these drives are not used in the experiments. The system runs on Ubuntu 18.04.5 LTS.

\Paragraph{Lower Query Cost} Figure \ref{fig:exp1} Parts (A)  measures latency for random negative queries (i.e., to non-existing keys). 
On each curve, one expansion occurs between two adjacent points, indicating a doubling of the data size. 
{\color{black} As the data  grows, the average latency  increases across all baselines as they outgrow the CPU caches.} Nevertheless, InfiniFilter's query cost deteriorates more rapidly since each query checks a growing number of hash tables. The FS method cannot expand indefinitely as eventually, all fingerprints run out of bits. {\color{black}Its performance is also more erratic as the filter's slot width changes, leading to cache misalignment.} In contrast, Aleph Filter supports unlimited expansions while maintaining stabler latency as each query checks at most one hash table. 

Part (B) measures the false positive rate (FPR) for the same queries as in Part (A). The FS method exhibits a skyrocketing FPR as fingerprints shrink across expansions. InfiniFilter and Aleph Filter  exhibit stabler FPRs that  match the model in Equation \ref{eq:FPR_fixed_width2}. 

{\color{black} Part (C) measures query latency for uniformly random existing keys. InfiniFilter is only slightly slower than Aleph Filter as most queries terminate after finding a matching entry in the main hash table. In contrast, Part (D) measures latency for queries targeting the oldest existing entries. InfiniFilter traverses multiple hash tables in this case, thus incurring significantly higher latency. Aleph Filter is  faster as each query returns a positive immediately after encountering a void entry in the main hash table. }


{
\color{blue}


}

\Paragraph{The Widening and Predictive Regimes} Figure \ref{fig:pred}  considers an application that requires an FPR of at most $\approx 1\%$ while expecting the data size to grow to $\approx 10^6$ entries. We initialize each baseline with the smallest memory footprint such that when the data size reaches $\approx 10^6$ entries, the FPR is at most $\approx 1\%$. InfiniFilter and Aleph Filter in the Widening Regime are each assigned 13 bits per entry. We initialize the FS method with 30 bits per entry and Aleph filter in the Predictive Regime with 22 bits per entry. We measure performance as the data size approaches and exceeds $\approx 10^6$ entries. 

Parts (B) and (C) show the takeaways. The FS method cannot meet the FPR requirement after the data size exceeds $\approx 10^6$ entries since all fingerprints shorten in each expansion.  In contrast,  InfiniFilter and Aleph Filter in the Widening Regime maintain a constant FPR from the get-go and even after the data size surpasses $\approx 10^6$ entries. The trade-off is a growing memory footprint. Aleph filter in the Predictive Regime also meets the FPR target while requiring a memory footprint on par with static filters when we meet the target data size. Even as the data outgrows our estimation, Aleph filter still requires less memory than in the Widening Regime as predicted by our model in Equations \ref{eq:predictive_FP2} and \ref{eq:predictive_FPR} and in Figure \ref{fig:predictive}. The trade-off is that it requires a few more bits per entry from the get-go, though this is a good deal: it is better to use more bits per entry when the data is small rather than large. 

For experimental control, Part (A) verifies that both variants of Aleph Filter exhibit the fastest queries, while Part (D) shows that all baselines have approximately the same insertion speed.



{ \color{black}
}

\begin{figure}[!t]
	\includegraphics[scale=0.36]{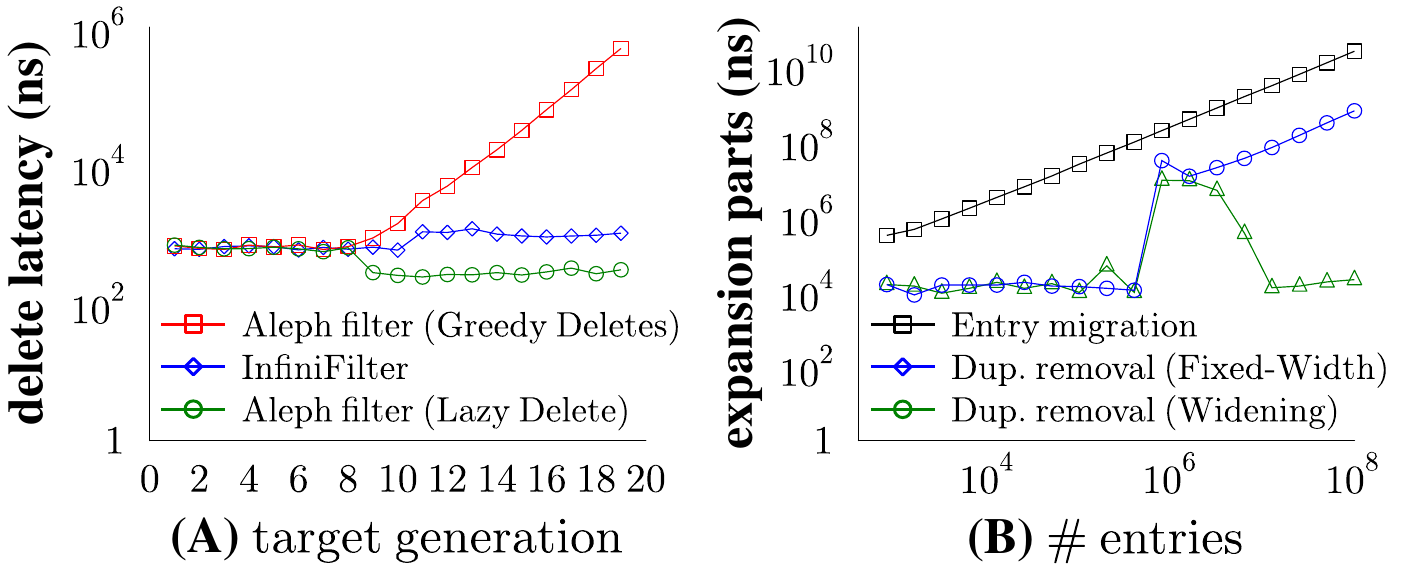}
	\vspace{-4mm}
	\caption{  Aleph Filter supports fast deletions by using tombstones and deferring the removal of void duplicates to the next expansion.	}
	\vspace{-2mm}
	\label{fig:deletes_eval}
\end{figure}

\Paragraph{Cheaper Deletes}  Figure \ref{fig:deletes_eval} Part (A) focuses on deletion latency. We show two variants of Aleph Filter with greedy vs. lazy deletes. The former identifies and removes void duplicates immediately during a deletion. The latter uses tombstones and removes void duplicates lazily during the next expansion as described in Section~\ref{sec:deletes}. We compare these baselines to InfiniFilter. All baselines are in the Fixed-Width Regime with 12-bit slots. 
We initialize each baseline with $2^9$ slots and perform insertions until it expands to $2^{25}$ slots. We then clone each baseline. For each clone, we measure latency for deleting $512$ random entries from the same generation (i.e., entries inserted in-between the same two expansions). As we move from left to right on the x-axis, we delete  entries from earlier generations (i.e., older entries). 

The figure shows that for all baselines, deletion latency is $\approx 600$ ns when deleting non-void entries (on the left-hand side). 
As we delete older entries, the latency for InfiniFilter increases to $\approx 800$ since more hash tables are accessed to find and remove them. For Aleph Filter with greedy deletes,  latency skyrockets as we target older entries since each entry has exponentially more duplicates in the main hash table, each of which must be removed. In contrast, for Aleph Filter with lazy deletes, latency decreases as we delete older entries since we only replace one void entry with a tombstone (i.e., we do not need to shift the remaining fingerprints in the cluster backward by one slot). Hence, Aleph filter's use of tombstones achieves constant time deletions. 
An experiment with rejuvenation operations yields a nearly identical figure, and so we omit it.


\Paragraph{Deletion Cost Gets Amortized} When Aleph filter deletes a void entry by replacing it with a tombstone, the removal of the entry's potentially many duplicates is deferred to the next expansion. To ensure that this does not degrade expansion speed, Figure \ref{fig:deletes_eval} Part~(B) measures this overhead in both the Fixed-Width and Widening Regimes against the cost of migrating entries into an expanded hash table. We run the experiment by inserting entries into an initially small empty filter, causing it to expand multiple times. From the $10^{\text{th}}$ expansion onward, we delete the oldest remaining generation of entries before the next expansion (e.g., After Expansion 15, we delete all entries inserted in Generation 5, etc.). During the next expansion, all void duplicates for this generation of entries are removed. This experiment represents the worst-case consistent toll that deletes can exact on expansions. 

In the Fixed-Width Regime, removing void duplicates takes approximately two orders of magnitude less time than the overhead of migrating entries into the expanded hash table. The reason is that the maximum number of void duplicates from each generation as a proportion of the filter size is small (i.e.,  $\approx 2^{-F}$ as shown in Section \ref{sec:analysis}). By contrast, in the Widening Regime, entries are slower to become void as they are created with increasing fingerprint lengths. Therefore, there are only a few generations of void entries to remove. Once they are removed, all subsequent deletions target entries in the main hash table, and so the removal of void duplicates inflicts no toll on expansion. The experiment demonstrates that in both regimes, the removal of void duplicates is heavily amortized with respect to the cost of expansion. Thus, insertions stay fast no matter how many deletes targeting old entries are in the workload. 

\begin{figure*}[!t]
	\includegraphics[scale=0.36]{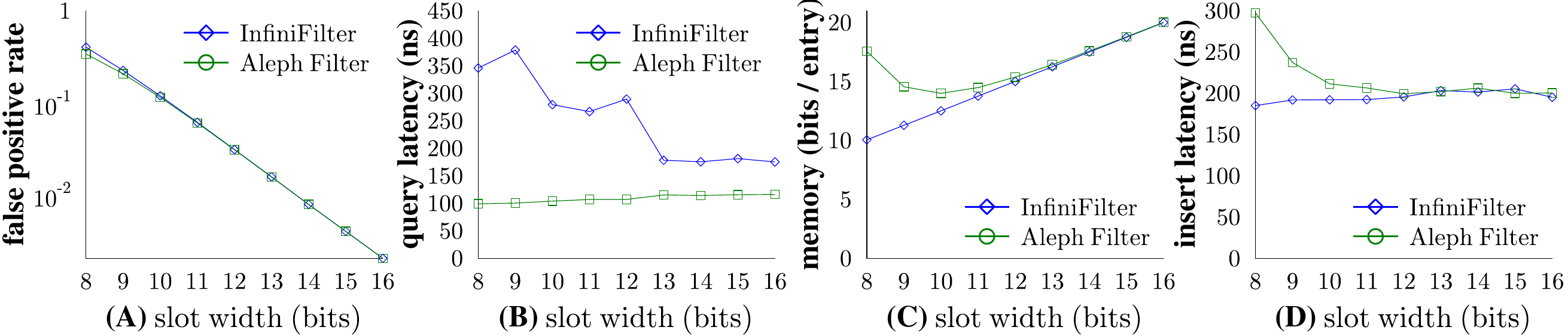}
	\vspace{-2mm}
	\caption{ \color{black} Aleph Filter improves on InfiniFilter across the board in terms of query cost, though its memory footprint and average insertion latency suffer when using a narrow slot width (i.e., fewer than 10 bits per entry).  }
	\vspace{-3mm}
	\label{fig:memory}
\end{figure*}

{ \color{black}
\Paragraph{Expansion Threshold} Figure \ref{fig:threshold} varies the threshold at which we expand InfiniFilter and Aleph Filter. A higher threshold makes these filters more compact and memory efficient, though it hurts performance as clusters in the underlying quotient filter become longer. Part (A) of the figure measures the average query latency right before each expansion as we insert $2^{26}$ keys. Each dot in this figure represents a trial starting with a small empty filter. With a higher threshold, InfiniFilter's query latency significantly increases as longer clusters must be traversed across multiple hash tables. In Aleph Filter, query latency increases more slowly as only one hash table is accessed per query. Part (B) shows that insertion performance is similar across these baselines as longer clusters need to be traversed before finding the target run for a new fingerprint.  
}

{ \color{black}
	\Paragraph{Transient Space Cost} Figure \ref{fig:memory_skew} Part (A) uses an independent thread to measure the filter's total memory footprint every two seconds as we insert $2^{29}$ keys. We observe significant memory spikes during expansion. 
	A spike begins whenever we allocate a larger main hash table and transfer fingerprints into it from the former main hash table. A spike ends when we deallocate the former main hash table. Handling expansions without such memory spikes is an intriguing future work direction. 
}

{ \color{black}
\Paragraph{Skew} Figure \ref{fig:memory_skew} Part (B) examines the impact of insertion skew. We insert keys from a Gaussian distribution with a mean of one billion and varying the standard deviation on the x-axis (from 1 million to 16 million) to control the likelihood of repeated insertions of the same key. Each dot represents a trial starting with a small empty filter followed by $2^{24}$ insertions. With both Aleph Filter and InfiniFilter, more skew (i.e., smaller standard deviation) leads to longer clusters and thus higher insertion latency. This problem can be alleviated by embedding a counter in the filter to count  repeating identical fingerprints rather than materializing each instance \cite{Pandey2017}, though this feature is not yet implemented in our library. 
}

{ \color{black}
\Paragraph{Slot Width} Figure \ref{fig:memory} compares Aleph Filter to InfiniFilter as we vary the initial slot width in the Fixed-Width Regime. 
Each dot represents a trial starting from a small empty filter and inserting $2^{26}$ keys. Part (A) shows  the FPR dropping for both baselines as the wider slots support longer fingerprints. Part (B) shows that with wider slots, InfiniFilter becomes more competitive with Aleph Filter in terms of query cost as there are fewer auxiliary hash tables to access along InfiniFilter's chain. Nevertheless, Aleph Filter still outperforms InfiniFilter across the board. Parts (C) and (D) highlight a limitation of Aleph Filter when operating with narrow slots (i.e., fewer than 10 bits per entry). In this case, the fraction of void entries in the filter becomes significant. This leads to premature expansion, which compromises the overall memory footprint as shown in Part~(C). This also degrades insertion throughput as the void entries entail more overhead to copy during expansion as shown in Part (D). The figure shows that it is best to operate Aleph Filter with 10 bits per entry or more. In this range, Aleph Filter provides a significant query cost improvement without a significant degradation to the overall memory footprint or insertion latency. In the real world, filters are typically allocated with 10-16 bits per entry,   making Aleph Filter viable across many applications  \cite{Dayan2021, Conway2023, Ren2017}. 
}

%% file: related-work.tex
\section{Related Work} \label{sec:related}

Some filter expansion approaches allocate additional empty filters into which more insertions can be made  \cite{Guo2006, Guo2009, Chen2017, Xie2007, Almeida2007, Williger2019}. The issue  is that multiple filters potentially need to be checked during a query or deletion. 
Other approaches \cite{Luo2019b, Wie2022} form a hash ring of buckets to support elastic expansion, yet this causes all operations to take $O(\lg N)$ time to search a binary tree for a given entry's bucket.
Yet another approach involves fetching a larger fingerprint for a key from storage, yet this  entails expensive I/Os {\cite{Wu2021}}. 
In contrast, Aleph filter provides constant time operations and does not require storage I/Os. 

Complementary approaches  expand at a finer granularity to prevent blocking regular operations and save space \cite{Wang2019, Wang2022, Wang2024}. 
Applying such techniques to Aleph filter could be impactful. 



Pagh, Segev, and Wieder prove a lower bound that if we initialize a filter to constant capacity and expand it to contain $N$ keys, the filter must at some point use at least $\lg \lg N$ bits per key in addition to what is required for a filter with a fixed capacity of $N$ keys~\cite{Pagh2013}. Aleph Filter's Widening and Predictive Regimes both meet this lower bound. The former meets it after the $N$ data entries are inserted, while the latter meets it before any entries are inserted. 

The idea of duplicating void entries within a filter hash table to support indefinite expansion with constant time queries can also be traced back to Pagh, Segev, and Wieder \cite{Pagh2013}. Taffy Cuckoo Filter~\cite{Apple2022} is an expandable Cuckoo filter \cite{Fan2014} based on these ideas, though it does not support deletes or rejuvenation operations.
The data structure by Liu, Yin, and Yu~\cite{Liu2020} performs constant time operations with high probability and a lower space overhead of $\lg \lg N + O(\lg \lg \lg  N)$ bits per entry if $N$ and $U$ are polynomially related. This matches the leading term of the lower bound. Nevertheless, this method has not been shown to support deletes, and it does not support unbounded growth beyond a universe size of $U$. 

Aleph filter improves on these works in several ways. (1) In the Widening Regime, it requires $O(\lg \lg N)$  rather than $O(\lg \lg U)$ bits per key to support unbounded growth while maintaining a stable FPR. This is a significant difference if $U$, the universe size, is unknown or significantly larger than $N$, the data size.  (2) To support deletes, the data structure in \cite{Pagh2013} keeps track of the age of entries using a binary age counter alongside each slot for $\lg \lg U$ additional bits per key. In contrast, Aleph filter keeps track of the ages of entries using a chain of exponentially smaller hash tables that take up negligible space. (3) The data structure in \cite{Pagh2013} handles fingerprint collisions by storing the full keys of entries with colliding fingerprints in an auxiliary dictionary. In contrast, Aleph filter can store multiple identical fingerprints in the main hash table  by employing Robin Hood hashing. This eliminates one dictionary access from the query path. (4) Aleph Filter introduces the Predictive Regime, which  significantly reduces the memory footprint.


%% file: conclusion.tex
\section{Conclusion}

We introduced Aleph Filter,  an infinitely expandable filter with constant time operations and improved FPR vs. memory trade-offs. 
Applying the Aleph Filter on top of other point filters \cite{Pandey2017, Bercea2019, Breslow2018, Graf2020, Even2022, Dillinger2022},  range filters  \cite{goswami2014, Zhang2018a, Luo2020x, Knorr2022, Vaidya2022, Mossner2023, Chen2024, Costa2024}, and adaptive filters  \cite{Mitzenmacher2018, Deeds2020, Kraska2018, Reviriego2023} can offer intriguing future work directions.



%% file: top.bbl

\begin{thebibliography}{63}


\ifx \showCODEN    \undefined \def \showCODEN     #1{\unskip}     \fi
\ifx \showDOI      \undefined \def \showDOI       #1{#1}\fi
\ifx \showISBNx    \undefined \def \showISBNx     #1{\unskip}     \fi
\ifx \showISBNxiii \undefined \def \showISBNxiii  #1{\unskip}     \fi
\ifx \showISSN     \undefined \def \showISSN      #1{\unskip}     \fi
\ifx \showLCCN     \undefined \def \showLCCN      #1{\unskip}     \fi
\ifx \shownote     \undefined \def \shownote      #1{#1}          \fi
\ifx \showarticletitle \undefined \def \showarticletitle #1{#1}   \fi
\ifx \showURL      \undefined \def \showURL       {\relax}        \fi
\providecommand\bibfield[2]{#2}
\providecommand\bibinfo[2]{#2}
\providecommand\natexlab[1]{#1}
\providecommand\showeprint[2][]{arXiv:#2}

\bibitem[\protect\citeauthoryear{Almeida, Baquero, Pregui{\c{c}}a, and
  Hutchison}{Almeida et~al\mbox{.}}{2007}]%
        {Almeida2007}
\bibfield{author}{\bibinfo{person}{Paulo~S{\'{e}}rgio Almeida},
  \bibinfo{person}{Carlos Baquero}, \bibinfo{person}{Nuno Pregui{\c{c}}a},
  {and} \bibinfo{person}{David Hutchison}.} \bibinfo{year}{2007}\natexlab{}.
\newblock \showarticletitle{{Scalable Bloom Filters}}.
\newblock \bibinfo{journal}{\emph{Inform. Process. Lett.}}
  (\bibinfo{year}{2007}).
\newblock


\bibitem[\protect\citeauthoryear{Andersen, Franklin, Kaminsky, Phanishayee,
  Tan, and Vasudevan}{Andersen et~al\mbox{.}}{2009}]%
        {Andersen2009}
\bibfield{author}{\bibinfo{person}{David~G. Andersen}, \bibinfo{person}{Jason
  Franklin}, \bibinfo{person}{Michael Kaminsky}, \bibinfo{person}{Amar
  Phanishayee}, \bibinfo{person}{Lawrence Tan}, {and} \bibinfo{person}{Vijay
  Vasudevan}.} \bibinfo{year}{2009}\natexlab{}.
\newblock \showarticletitle{{FAWN: A Fast Array of Wimpy Nodes}}.
\newblock \bibinfo{journal}{\emph{SOSP}} (\bibinfo{year}{2009}).
\newblock


\bibitem[\protect\citeauthoryear{Apple}{Apple}{2022}]%
        {Apple2022}
\bibfield{author}{\bibinfo{person}{Jim Apple}.}
  \bibinfo{year}{2022}\natexlab{}.
\newblock \showarticletitle{Stretching your data with taffy filters}.
\newblock \bibinfo{journal}{\emph{Software: Practice and Experience}}
  (\bibinfo{year}{2022}).
\newblock


\bibitem[\protect\citeauthoryear{Bender, Farach-Colton, Johnson, Kraner,
  Kuszmaul, Medjedovic, Montes, Shetty, Spillane, and Zadok}{Bender
  et~al\mbox{.}}{2012}]%
        {Bender2012}
\bibfield{author}{\bibinfo{person}{Michael~A. Bender}, \bibinfo{person}{Martin
  Farach-Colton}, \bibinfo{person}{Rob Johnson}, \bibinfo{person}{Russell
  Kraner}, \bibinfo{person}{Bradley~C. Kuszmaul}, \bibinfo{person}{Dzejla
  Medjedovic}, \bibinfo{person}{Pablo Montes}, \bibinfo{person}{Pradeep
  Shetty}, \bibinfo{person}{Richard~P. Spillane}, {and} \bibinfo{person}{Erez
  Zadok}.} \bibinfo{year}{2012}\natexlab{}.
\newblock \showarticletitle{{Don't Thrash: How to Cache Your Hash on Flash}}.
\newblock \bibinfo{journal}{\emph{PVLDB}} (\bibinfo{year}{2012}).
\newblock


\bibitem[\protect\citeauthoryear{Bercea and Even}{Bercea and Even}{2020}]%
        {Bercea2019}
\bibfield{author}{\bibinfo{person}{Ioana~O Bercea} {and} \bibinfo{person}{Guy
  Even}.} \bibinfo{year}{2020}\natexlab{}.
\newblock \showarticletitle{Fully-Dynamic Space-Efficient Dictionaries and
  Filters with Constant Number of Memory Accesses}.
\newblock \bibinfo{journal}{\emph{SWAT}} (\bibinfo{year}{2020}).
\newblock


\bibitem[\protect\citeauthoryear{Bloom}{Bloom}{1970}]%
        {Bloom1970}
\bibfield{author}{\bibinfo{person}{Burton~H. Bloom}.}
  \bibinfo{year}{1970}\natexlab{}.
\newblock \showarticletitle{{Space/Time Trade-offs in Hash Coding with
  Allowable Errors}}.
\newblock \bibinfo{journal}{\emph{CACM}} (\bibinfo{year}{1970}).
\newblock


\bibitem[\protect\citeauthoryear{Breslow and Jayasena}{Breslow and
  Jayasena}{2018}]%
        {Breslow2018}
\bibfield{author}{\bibinfo{person}{Alex~D Breslow} {and}
  \bibinfo{person}{Nuwan~S Jayasena}.} \bibinfo{year}{2018}\natexlab{}.
\newblock \showarticletitle{Morton Filters: Faster, Space-Efficient Cuckoo
  Filters via Biasing, compression, and decoupled logical sparsity}.
\newblock \bibinfo{journal}{\emph{PVLDB}} (\bibinfo{year}{2018}).
\newblock


\bibitem[\protect\citeauthoryear{Broder and Mitzenmacher}{Broder and
  Mitzenmacher}{2002}]%
        {Broder2002}
\bibfield{author}{\bibinfo{person}{Andrei~Z. Broder} {and}
  \bibinfo{person}{Michael Mitzenmacher}.} \bibinfo{year}{2002}\natexlab{}.
\newblock \showarticletitle{{Network Applications of Bloom Filters: A Survey}}.
\newblock \bibinfo{journal}{\emph{Internet Mathematics}}  \bibinfo{volume}{1}
  (\bibinfo{year}{2002}), \bibinfo{pages}{636--646}.
\newblock


\bibitem[\protect\citeauthoryear{Carter, Floyd, Gill, Markowsky, and
  Wegman}{Carter et~al\mbox{.}}{1978}]%
        {Carter1978}
\bibfield{author}{\bibinfo{person}{Larry Carter}, \bibinfo{person}{Robert
  Floyd}, \bibinfo{person}{John Gill}, \bibinfo{person}{George Markowsky},
  {and} \bibinfo{person}{Mark Wegman}.} \bibinfo{year}{1978}\natexlab{}.
\newblock \showarticletitle{Exact and Approximate Membership Testers}. In
  \bibinfo{booktitle}{\emph{STOC}}.
\newblock


\bibitem[\protect\citeauthoryear{Celis, Larson, and Munro}{Celis
  et~al\mbox{.}}{1985}]%
        {Celis1985}
\bibfield{author}{\bibinfo{person}{Pedro Celis}, \bibinfo{person}{Per-Ake
  Larson}, {and} \bibinfo{person}{J~Ian Munro}.}
  \bibinfo{year}{1985}\natexlab{}.
\newblock \showarticletitle{Robin Hood Hashing}.
\newblock \bibinfo{journal}{\emph{FOCS}} (\bibinfo{year}{1985}).
\newblock


\bibitem[\protect\citeauthoryear{Chandramouli, Prasaad, Kossmann, Levandoski,
  Hunter, and Barnett}{Chandramouli et~al\mbox{.}}{2018}]%
        {Chandramouli2018}
\bibfield{author}{\bibinfo{person}{Badrish Chandramouli}, \bibinfo{person}{Guna
  Prasaad}, \bibinfo{person}{Donald Kossmann}, \bibinfo{person}{Justin~J
  Levandoski}, \bibinfo{person}{James Hunter}, {and} \bibinfo{person}{Mike
  Barnett}.} \bibinfo{year}{2018}\natexlab{}.
\newblock \showarticletitle{{FASTER: A Concurrent Key-Value Store with In-Place
  Updates}}.
\newblock \bibinfo{journal}{\emph{SIGMOD}} (\bibinfo{year}{2018}).
\newblock


\bibitem[\protect\citeauthoryear{Chen, He, Li, and Luo}{Chen
  et~al\mbox{.}}{2024}]%
        {Chen2024}
\bibfield{author}{\bibinfo{person}{Guanduo Chen}, \bibinfo{person}{Zhenying
  He}, \bibinfo{person}{Meng Li}, {and} \bibinfo{person}{Siqiang Luo}.}
  \bibinfo{year}{2024}\natexlab{}.
\newblock \showarticletitle{Oasis: An Optimal Disjoint Segmented Learned Range
  Filter}.
\newblock \bibinfo{journal}{\emph{PVLDB}} (\bibinfo{year}{2024}).
\newblock


\bibitem[\protect\citeauthoryear{Chen, Liao, Jin, and Wu}{Chen
  et~al\mbox{.}}{2017}]%
        {Chen2017}
\bibfield{author}{\bibinfo{person}{Hanhua Chen}, \bibinfo{person}{Liangyi
  Liao}, \bibinfo{person}{Hai Jin}, {and} \bibinfo{person}{Jie Wu}.}
  \bibinfo{year}{2017}\natexlab{}.
\newblock \showarticletitle{The Dynamic Cuckoo Filter}.
\newblock \bibinfo{journal}{\emph{ICNP}} (\bibinfo{year}{2017}).
\newblock


\bibitem[\protect\citeauthoryear{Clerry}{Clerry}{1984}]%
        {Clerry1984}
\bibfield{author}{\bibinfo{person}{John~G. Clerry}.}
  \bibinfo{year}{1984}\natexlab{}.
\newblock \showarticletitle{Compact hash tables using bidirectional linear
  probing}.
\newblock \bibinfo{journal}{\emph{IEEE Trans. Comput.}} (\bibinfo{year}{1984}).
\newblock


\bibitem[\protect\citeauthoryear{Collet}{Collet}{2023}]%
        {XXHash}
\bibfield{author}{\bibinfo{person}{Yann Collet}.}
  \bibinfo{year}{2023}\natexlab{}.
\newblock \showarticletitle{{XXHash}}.
\newblock \bibinfo{journal}{\emph{https://github.com/Cyan4973/xxHash}}
  (\bibinfo{year}{2023}).
\newblock


\bibitem[\protect\citeauthoryear{Conway, Farach-Colton, and Johnson}{Conway
  et~al\mbox{.}}{2023}]%
        {Conway2023}
\bibfield{author}{\bibinfo{person}{Alex Conway}, \bibinfo{person}{Mart{\'\i}n
  Farach-Colton}, {and} \bibinfo{person}{Rob Johnson}.}
  \bibinfo{year}{2023}\natexlab{}.
\newblock \showarticletitle{SplinterDB and Maplets: Improving the Tradeoffs in
  Key-Value Store Compaction Policy}.
\newblock \bibinfo{journal}{\emph{SIGMOD}} (\bibinfo{year}{2023}).
\newblock


\bibitem[\protect\citeauthoryear{Costa, Ferragina, and Vinciguerra}{Costa
  et~al\mbox{.}}{2024}]%
        {Costa2024}
\bibfield{author}{\bibinfo{person}{Marco Costa}, \bibinfo{person}{Paolo
  Ferragina}, {and} \bibinfo{person}{Giorgio Vinciguerra}.}
  \bibinfo{year}{2024}\natexlab{}.
\newblock \showarticletitle{Grafite: Taming Adversarial Queries with Optimal
  Range Filters}.
\newblock \bibinfo{journal}{\emph{SIGMOD}} (\bibinfo{year}{2024}).
\newblock


\bibitem[\protect\citeauthoryear{Dayan, Athanassoulis, and Idreos}{Dayan
  et~al\mbox{.}}{2017}]%
        {Dayan2017}
\bibfield{author}{\bibinfo{person}{Niv Dayan}, \bibinfo{person}{Manos
  Athanassoulis}, {and} \bibinfo{person}{Stratos Idreos}.}
  \bibinfo{year}{2017}\natexlab{}.
\newblock \showarticletitle{{Monkey: Optimal Navigable Key-Value Store}}.
\newblock \bibinfo{journal}{\emph{SIGMOD}} (\bibinfo{year}{2017}).
\newblock


\bibitem[\protect\citeauthoryear{Dayan, Bercea, Reviriego, and Pagh}{Dayan
  et~al\mbox{.}}{2023}]%
        {Dayan2023}
\bibfield{author}{\bibinfo{person}{Niv Dayan}, \bibinfo{person}{Ioana Bercea},
  \bibinfo{person}{Pedro Reviriego}, {and} \bibinfo{person}{Rasmus Pagh}.}
  \bibinfo{year}{2023}\natexlab{}.
\newblock \showarticletitle{InfiniFilter: Expanding Filters to Infinity and
  Beyond}.
\newblock \bibinfo{journal}{\emph{SIGMOD}} (\bibinfo{year}{2023}).
\newblock


\bibitem[\protect\citeauthoryear{Dayan and Idreos}{Dayan and Idreos}{2018}]%
        {Dayan2018}
\bibfield{author}{\bibinfo{person}{Niv Dayan} {and} \bibinfo{person}{Stratos
  Idreos}.} \bibinfo{year}{2018}\natexlab{}.
\newblock \showarticletitle{{Dostoevsky: Better Space-Time Trade-Offs for
  LSM-Tree Based Key-Value Stores via Adaptive Removal of Superfluous
  Merging}}.
\newblock \bibinfo{journal}{\emph{SIGMOD}} (\bibinfo{year}{2018}).
\newblock


\bibitem[\protect\citeauthoryear{Dayan and Idreos}{Dayan and Idreos}{2019}]%
        {Dayan2019}
\bibfield{author}{\bibinfo{person}{Niv Dayan} {and} \bibinfo{person}{Stratos
  Idreos}.} \bibinfo{year}{2019}\natexlab{}.
\newblock \showarticletitle{The Log-Structured Merge-Bush \& the Wacky
  Continuum}.
\newblock \bibinfo{journal}{\emph{SIGMOD}} (\bibinfo{year}{2019}).
\newblock


\bibitem[\protect\citeauthoryear{Dayan and Twitto}{Dayan and Twitto}{2021}]%
        {Dayan2021}
\bibfield{author}{\bibinfo{person}{Niv Dayan} {and} \bibinfo{person}{Moshe
  Twitto}.} \bibinfo{year}{2021}\natexlab{}.
\newblock \showarticletitle{Chucky: A Succinct Cuckoo Filter for LSM-Tree}.
\newblock \bibinfo{journal}{\emph{SIGMOD}} (\bibinfo{year}{2021}).
\newblock


\bibitem[\protect\citeauthoryear{Dayan, Twitto, Rochman, Beitler, Zion,
  Bortnikov, Dashevsky, Frishman, Ginzburg, Maly, et~al\mbox{.}}{Dayan
  et~al\mbox{.}}{2021}]%
        {Dayan2021B}
\bibfield{author}{\bibinfo{person}{Niv Dayan}, \bibinfo{person}{Moshe Twitto},
  \bibinfo{person}{Yuval Rochman}, \bibinfo{person}{Uri Beitler},
  \bibinfo{person}{Itai~Ben Zion}, \bibinfo{person}{Edward Bortnikov},
  \bibinfo{person}{Shmuel Dashevsky}, \bibinfo{person}{Ofer Frishman},
  \bibinfo{person}{Evgeni Ginzburg}, \bibinfo{person}{Igal Maly},
  {et~al\mbox{.}}} \bibinfo{year}{2021}\natexlab{}.
\newblock \showarticletitle{The End of Moore's Law and the Rise of the Data
  Processor}.
\newblock \bibinfo{journal}{\emph{PVLDB}} (\bibinfo{year}{2021}).
\newblock


\bibitem[\protect\citeauthoryear{Debnath, Sengupta, and Li}{Debnath
  et~al\mbox{.}}{2010}]%
        {Debnath2010}
\bibfield{author}{\bibinfo{person}{Biplob Debnath}, \bibinfo{person}{Sudipta
  Sengupta}, {and} \bibinfo{person}{Jin Li}.} \bibinfo{year}{2010}\natexlab{}.
\newblock \showarticletitle{{FlashStore: High Throughput Persistent Key-Value
  Store}}.
\newblock \bibinfo{journal}{\emph{PVLDB}} (\bibinfo{year}{2010}).
\newblock


\bibitem[\protect\citeauthoryear{Debnath, Sengupta, and Li}{Debnath
  et~al\mbox{.}}{2011}]%
        {Debnath2011}
\bibfield{author}{\bibinfo{person}{Biplob Debnath}, \bibinfo{person}{Sudipta
  Sengupta}, {and} \bibinfo{person}{Jin Li}.} \bibinfo{year}{2011}\natexlab{}.
\newblock \showarticletitle{{SkimpyStash: RAM space skimpy key-value store on
  flash-based storage}}.
\newblock \bibinfo{journal}{\emph{SIGMOD}} (\bibinfo{year}{2011}).
\newblock


\bibitem[\protect\citeauthoryear{Deeds, Hentschel, and Idreos}{Deeds
  et~al\mbox{.}}{2020}]%
        {Deeds2020}
\bibfield{author}{\bibinfo{person}{Kyle Deeds}, \bibinfo{person}{Brian
  Hentschel}, {and} \bibinfo{person}{Stratos Idreos}.}
  \bibinfo{year}{2020}\natexlab{}.
\newblock \showarticletitle{Stacked filters: learning to filter by structure}.
\newblock \bibinfo{journal}{\emph{PVLDB}} (\bibinfo{year}{2020}).
\newblock


\bibitem[\protect\citeauthoryear{Dillinger, H{\"u}bschle-Schneider, Sanders,
  and Walzer}{Dillinger et~al\mbox{.}}{2022}]%
        {Dillinger2022}
\bibfield{author}{\bibinfo{person}{Peter~C Dillinger}, \bibinfo{person}{Lorenz
  H{\"u}bschle-Schneider}, \bibinfo{person}{Peter Sanders}, {and}
  \bibinfo{person}{Stefan Walzer}.} \bibinfo{year}{2022}\natexlab{}.
\newblock \showarticletitle{Fast Succinct Retrieval and Approximate Membership
  Using Ribbon}.
\newblock \bibinfo{journal}{\emph{SEA}} (\bibinfo{year}{2022}).
\newblock


\bibitem[\protect\citeauthoryear{Dillinger and Manolios}{Dillinger and
  Manolios}{2009}]%
        {Dillinger2009}
\bibfield{author}{\bibinfo{person}{Peter~C. Dillinger} {and}
  \bibinfo{person}{Panagiotis~Pete Manolios}.} \bibinfo{year}{2009}\natexlab{}.
\newblock \showarticletitle{Fast, All-Purpose State Storage}.
\newblock \bibinfo{journal}{\emph{SPIN}} (\bibinfo{year}{2009}).
\newblock


\bibitem[\protect\citeauthoryear{Even, Even, and Morrison}{Even
  et~al\mbox{.}}{2022}]%
        {Even2022}
\bibfield{author}{\bibinfo{person}{Tomer Even}, \bibinfo{person}{Guy Even},
  {and} \bibinfo{person}{Adam Morrison}.} \bibinfo{year}{2022}\natexlab{}.
\newblock \showarticletitle{{Prefix Filter: Practically and Theoretically
  Better Than Bloom}}.
\newblock \bibinfo{journal}{\emph{PVLDB}} (\bibinfo{year}{2022}).
\newblock


\bibitem[\protect\citeauthoryear{Fan, Andersen, Kaminsky, and Mitzenmacher}{Fan
  et~al\mbox{.}}{2014}]%
        {Fan2014}
\bibfield{author}{\bibinfo{person}{Bin Fan}, \bibinfo{person}{David~G.
  Andersen}, \bibinfo{person}{Michael Kaminsky}, {and} \bibinfo{person}{Michael
  Mitzenmacher}.} \bibinfo{year}{2014}\natexlab{}.
\newblock \showarticletitle{{Cuckoo Filter: Practically Better Than Bloom}}.
\newblock \bibinfo{journal}{\emph{CoNEXT}} (\bibinfo{year}{2014}).
\newblock


\bibitem[\protect\citeauthoryear{Goswami, Gr{\o}nlund, Larsen, and
  Pagh}{Goswami et~al\mbox{.}}{2014}]%
        {goswami2014}
\bibfield{author}{\bibinfo{person}{Mayank Goswami}, \bibinfo{person}{Allan
  Gr{\o}nlund}, \bibinfo{person}{Kasper~Green Larsen}, {and}
  \bibinfo{person}{Rasmus Pagh}.} \bibinfo{year}{2014}\natexlab{}.
\newblock \showarticletitle{Approximate Range Emptiness in Constant Time and
  Optimal Space}.
\newblock \bibinfo{journal}{\emph{SODA}} (\bibinfo{year}{2014}).
\newblock


\bibitem[\protect\citeauthoryear{Graf and Lemire}{Graf and Lemire}{2020}]%
        {Graf2020}
\bibfield{author}{\bibinfo{person}{Thomas~Mueller Graf} {and}
  \bibinfo{person}{Daniel Lemire}.} \bibinfo{year}{2020}\natexlab{}.
\newblock \showarticletitle{Xor Filters: Faster and Smaller Than Bloom and
  Cuckoo Filters}.
\newblock \bibinfo{journal}{\emph{JEA}} (\bibinfo{year}{2020}).
\newblock


\bibitem[\protect\citeauthoryear{Guo, Wu, Chen, and Luo}{Guo
  et~al\mbox{.}}{2006}]%
        {Guo2006}
\bibfield{author}{\bibinfo{person}{Deke Guo}, \bibinfo{person}{Jie Wu},
  \bibinfo{person}{Honghui Chen}, {and} \bibinfo{person}{Xueshan Luo}.}
  \bibinfo{year}{2006}\natexlab{}.
\newblock \showarticletitle{Theory and Network Applications of Dynamic Bloom
  Filters}.
\newblock \bibinfo{journal}{\emph{INFOCOM}} (\bibinfo{year}{2006}).
\newblock


\bibitem[\protect\citeauthoryear{Guo, Wu, Chen, Yuan, and Luo}{Guo
  et~al\mbox{.}}{2009}]%
        {Guo2009}
\bibfield{author}{\bibinfo{person}{Deke Guo}, \bibinfo{person}{Jie Wu},
  \bibinfo{person}{Honghui Chen}, \bibinfo{person}{Ye Yuan}, {and}
  \bibinfo{person}{Xueshan Luo}.} \bibinfo{year}{2009}\natexlab{}.
\newblock \showarticletitle{The Dynamic Bloom Filters}.
\newblock \bibinfo{journal}{\emph{IEEE Trans Knowl Data Eng}}
  (\bibinfo{year}{2009}).
\newblock


\bibitem[\protect\citeauthoryear{Idreos, Dayan, Qin, Akmanalp, Hilgard, Ross,
  Lennon, Jain, Gupta, Li, and Zhu}{Idreos et~al\mbox{.}}{2019}]%
        {Idreos2019}
\bibfield{author}{\bibinfo{person}{Stratos Idreos}, \bibinfo{person}{Niv
  Dayan}, \bibinfo{person}{Wilson Qin}, \bibinfo{person}{Mali Akmanalp},
  \bibinfo{person}{Sophie Hilgard}, \bibinfo{person}{Andrew Ross},
  \bibinfo{person}{James Lennon}, \bibinfo{person}{Varun Jain},
  \bibinfo{person}{Harshita Gupta}, \bibinfo{person}{David Li}, {and}
  \bibinfo{person}{Zichen Zhu}.} \bibinfo{year}{2019}\natexlab{}.
\newblock \showarticletitle{Design Continuums and the Path Toward
  Self-Designing Key-Value Stores that Know and Learn}. In
  \bibinfo{booktitle}{\emph{CIDR}}.
\newblock


\bibitem[\protect\citeauthoryear{Knorr, Lemaire, Lim, Luo, Zhang, Idreos, and
  Mitzenmacher}{Knorr et~al\mbox{.}}{2022}]%
        {Knorr2022}
\bibfield{author}{\bibinfo{person}{Eric~R Knorr}, \bibinfo{person}{Baptiste
  Lemaire}, \bibinfo{person}{Andrew Lim}, \bibinfo{person}{Siqiang Luo},
  \bibinfo{person}{Huanchen Zhang}, \bibinfo{person}{Stratos Idreos}, {and}
  \bibinfo{person}{Michael Mitzenmacher}.} \bibinfo{year}{2022}\natexlab{}.
\newblock \showarticletitle{Proteus: A Self-Designing Range Filter}.
\newblock \bibinfo{journal}{\emph{SIGMOD}} (\bibinfo{year}{2022}).
\newblock


\bibitem[\protect\citeauthoryear{Kraska, Beutel, Chi, Dean, and
  Polyzotis}{Kraska et~al\mbox{.}}{2018}]%
        {Kraska2018}
\bibfield{author}{\bibinfo{person}{Tim Kraska}, \bibinfo{person}{Alex Beutel},
  \bibinfo{person}{Ed~H Chi}, \bibinfo{person}{Jeffrey Dean}, {and}
  \bibinfo{person}{Neoklis Polyzotis}.} \bibinfo{year}{2018}\natexlab{}.
\newblock \showarticletitle{{The Case for Learned Index Structures}}.
\newblock \bibinfo{journal}{\emph{SIGMOD}} (\bibinfo{year}{2018}).
\newblock


\bibitem[\protect\citeauthoryear{Liu, Yin, and Yu}{Liu et~al\mbox{.}}{2020}]%
        {Liu2020}
\bibfield{author}{\bibinfo{person}{Mingmou Liu}, \bibinfo{person}{Yitong Yin},
  {and} \bibinfo{person}{Huacheng Yu}.} \bibinfo{year}{2020}\natexlab{}.
\newblock \showarticletitle{Succinct Filters for Sets of Unknown Sizes}.
\newblock \bibinfo{journal}{\emph{{ICALP}}} (\bibinfo{year}{2020}).
\newblock


\bibitem[\protect\citeauthoryear{Luo, Guo, Rottenstreich, Ma, Luo, and Ren}{Luo
  et~al\mbox{.}}{2019}]%
        {Luo2019b}
\bibfield{author}{\bibinfo{person}{Lailong Luo}, \bibinfo{person}{Deke Guo},
  \bibinfo{person}{Ori Rottenstreich}, \bibinfo{person}{Richard~TB Ma},
  \bibinfo{person}{Xueshan Luo}, {and} \bibinfo{person}{Bangbang Ren}.}
  \bibinfo{year}{2019}\natexlab{}.
\newblock \showarticletitle{The Consistent Cuckoo Filter}.
\newblock \bibinfo{journal}{\emph{INFOCOM}} (\bibinfo{year}{2019}).
\newblock


\bibitem[\protect\citeauthoryear{Luo, Chatterjee, Ketsetsidis, Dayan, Qin, and
  Idreos}{Luo et~al\mbox{.}}{2020}]%
        {Luo2020x}
\bibfield{author}{\bibinfo{person}{Siqiang Luo}, \bibinfo{person}{Subarna
  Chatterjee}, \bibinfo{person}{Rafael Ketsetsidis}, \bibinfo{person}{Niv
  Dayan}, \bibinfo{person}{Wilson Qin}, {and} \bibinfo{person}{Stratos
  Idreos}.} \bibinfo{year}{2020}\natexlab{}.
\newblock \showarticletitle{Rosetta: A Robust Space-Time Optimized Range Filter
  for Key-Value Stores}.
\newblock \bibinfo{journal}{\emph{SIGMOD}} (\bibinfo{year}{2020}).
\newblock


\bibitem[\protect\citeauthoryear{Mitzenmacher, Pontarelli, and
  Reviriego}{Mitzenmacher et~al\mbox{.}}{2018}]%
        {Mitzenmacher2018}
\bibfield{author}{\bibinfo{person}{Michael Mitzenmacher},
  \bibinfo{person}{Salvatore Pontarelli}, {and} \bibinfo{person}{Pedro
  Reviriego}.} \bibinfo{year}{2018}\natexlab{}.
\newblock \showarticletitle{Adaptive cuckoo filters}.
\newblock \bibinfo{journal}{\emph{SIAM ALENEX}} (\bibinfo{year}{2018}).
\newblock


\bibitem[\protect\citeauthoryear{M{\"o}{\ss}ner, Riegger, Bernhardt, and
  Petrov}{M{\"o}{\ss}ner et~al\mbox{.}}{2022}]%
        {Mossner2023}
\bibfield{author}{\bibinfo{person}{Bernhard M{\"o}{\ss}ner},
  \bibinfo{person}{Christian Riegger}, \bibinfo{person}{Arthur Bernhardt},
  {and} \bibinfo{person}{Ilia Petrov}.} \bibinfo{year}{2022}\natexlab{}.
\newblock \showarticletitle{bloomRF: On performing range-queries in
  Bloom-Filters with piecewise-monotone hash functions and prefix hashing}.
\newblock \bibinfo{journal}{\emph{EDBT}} (\bibinfo{year}{2022}).
\newblock


\bibitem[\protect\citeauthoryear{Pagh, Segev, and Wieder}{Pagh
  et~al\mbox{.}}{2013}]%
        {Pagh2013}
\bibfield{author}{\bibinfo{person}{Rasmus Pagh}, \bibinfo{person}{Gil Segev},
  {and} \bibinfo{person}{Udi Wieder}.} \bibinfo{year}{2013}\natexlab{}.
\newblock \showarticletitle{How to Approximate a Set Without Knowing its Size
  in Advance}.
\newblock \bibinfo{journal}{\emph{FOCS}} (\bibinfo{year}{2013}).
\newblock


\bibitem[\protect\citeauthoryear{Pandey, Bender, Johnson, and Patro}{Pandey
  et~al\mbox{.}}{2017}]%
        {Pandey2017}
\bibfield{author}{\bibinfo{person}{Prashant Pandey}, \bibinfo{person}{Michael~A
  Bender}, \bibinfo{person}{Rob Johnson}, {and} \bibinfo{person}{Rob Patro}.}
  \bibinfo{year}{2017}\natexlab{}.
\newblock \showarticletitle{A General-Purpose Counting Filter: Making Every Bit
  Count}.
\newblock \bibinfo{journal}{\emph{SIGMOD}} (\bibinfo{year}{2017}).
\newblock


\bibitem[\protect\citeauthoryear{Pandey, Conway, Durie, Bender, Farach-Colton,
  and Johnson}{Pandey et~al\mbox{.}}{2021}]%
        {Pandey2021}
\bibfield{author}{\bibinfo{person}{Prashant Pandey}, \bibinfo{person}{Alex
  Conway}, \bibinfo{person}{Joe Durie}, \bibinfo{person}{Michael~A Bender},
  \bibinfo{person}{Martin Farach-Colton}, {and} \bibinfo{person}{Rob Johnson}.}
  \bibinfo{year}{2021}\natexlab{}.
\newblock \showarticletitle{Vector Quotient Filters: Overcoming the Time/Space
  Trade-Off in Filter Design}. In \bibinfo{booktitle}{\emph{SIGMOD}}.
\newblock


\bibitem[\protect\citeauthoryear{Pandey, Farach-Colton, Dayan, and
  Zhang}{Pandey et~al\mbox{.}}{2024}]%
        {Pandey2024}
\bibfield{author}{\bibinfo{person}{Prashant Pandey},
  \bibinfo{person}{Mart{\'\i}n Farach-Colton}, \bibinfo{person}{Niv Dayan},
  {and} \bibinfo{person}{Huanchen Zhang}.} \bibinfo{year}{2024}\natexlab{}.
\newblock \showarticletitle{Beyond Bloom: A Tutorial on Future Feature-Rich
  Filters}.
\newblock \bibinfo{journal}{\emph{SIGMOD}} (\bibinfo{year}{2024}).
\newblock


\bibitem[\protect\citeauthoryear{Peterson}{Peterson}{1957}]%
        {Peterson1957}
\bibfield{author}{\bibinfo{person}{W~Wesley Peterson}.}
  \bibinfo{year}{1957}\natexlab{}.
\newblock \showarticletitle{Addressing for random-access storage}.
\newblock \bibinfo{journal}{\emph{IBM journal of Research and Development}}
  (\bibinfo{year}{1957}).
\newblock


\bibitem[\protect\citeauthoryear{Ren, Zheng, Arulraj, and Gibson}{Ren
  et~al\mbox{.}}{2017}]%
        {Ren2017}
\bibfield{author}{\bibinfo{person}{Kai Ren}, \bibinfo{person}{Qing Zheng},
  \bibinfo{person}{Joy Arulraj}, {and} \bibinfo{person}{Garth Gibson}.}
  \bibinfo{year}{2017}\natexlab{}.
\newblock \showarticletitle{{SlimDB: A Space-Efficient Key-Value Storage Engine
  For Semi-Sorted Data}}.
\newblock \bibinfo{journal}{\emph{PVLDB}} (\bibinfo{year}{2017}).
\newblock


\bibitem[\protect\citeauthoryear{Reviriego, Apple, Alonso, Ertl, and
  Dayan}{Reviriego et~al\mbox{.}}{2023}]%
        {Reviriego2023}
\bibfield{author}{\bibinfo{person}{Pedro Reviriego}, \bibinfo{person}{Jim
  Apple}, \bibinfo{person}{Alvaro Alonso}, \bibinfo{person}{Otmar Ertl}, {and}
  \bibinfo{person}{Niv Dayan}.} \bibinfo{year}{2023}\natexlab{}.
\newblock \showarticletitle{Cardinality Estimation Adaptive Cuckoo Filters
  (CE-ACF): Approximate Membership Check and Distinct Query Count for
  High-Speed Network Monitoring}.
\newblock \bibinfo{journal}{\emph{IEEE/ACM Trans. Netw.}}
  (\bibinfo{year}{2023}).
\newblock


\bibitem[\protect\citeauthoryear{Reviriego, Gonz{\'a}lez, Dayan, Huecas, Liu,
  and Lombardi}{Reviriego et~al\mbox{.}}{2024}]%
        {Reviriego2024}
\bibfield{author}{\bibinfo{person}{Pedro Reviriego}, \bibinfo{person}{Miguel
  Gonz{\'a}lez}, \bibinfo{person}{Niv Dayan}, \bibinfo{person}{Gabriel Huecas},
  \bibinfo{person}{Shanshan Liu}, {and} \bibinfo{person}{Fabrizio Lombardi}.}
  \bibinfo{year}{2024}\natexlab{}.
\newblock \showarticletitle{On the Security of Quotient Filters: Attacks and
  Potential Countermeasures}.
\newblock \bibinfo{journal}{\emph{IEEE Trans. Comput.}} (\bibinfo{year}{2024}).
\newblock


\bibitem[\protect\citeauthoryear{Sarkar, Dayan, and Athanassoulis}{Sarkar
  et~al\mbox{.}}{2023}]%
        {Sarkar2023}
\bibfield{author}{\bibinfo{person}{Subhadeep Sarkar}, \bibinfo{person}{Niv
  Dayan}, {and} \bibinfo{person}{Manos Athanassoulis}.}
  \bibinfo{year}{2023}\natexlab{}.
\newblock \showarticletitle{{The LSM Design Space and its Read Optimizations}}.
\newblock \bibinfo{journal}{\emph{ICDE}} (\bibinfo{year}{2023}).
\newblock


\bibitem[\protect\citeauthoryear{Tarkoma, Rothenberg, and Lagerspetz}{Tarkoma
  et~al\mbox{.}}{2012}]%
        {Tarkoma2012}
\bibfield{author}{\bibinfo{person}{Sasu Tarkoma},
  \bibinfo{person}{Christian~Esteve Rothenberg}, {and} \bibinfo{person}{Eemil
  Lagerspetz}.} \bibinfo{year}{2012}\natexlab{}.
\newblock \showarticletitle{{Theory and Practice of Bloom Filters for
  Distributed Systems}}.
\newblock \bibinfo{journal}{\emph{IEEE Commun. Surv. Tutor}}
  (\bibinfo{year}{2012}).
\newblock


\bibitem[\protect\citeauthoryear{Vaidya, Chatterjee, Knorr, Mitzenmacher,
  Idreos, and Kraska}{Vaidya et~al\mbox{.}}{2022}]%
        {Vaidya2022}
\bibfield{author}{\bibinfo{person}{Kapil Vaidya}, \bibinfo{person}{Subarna
  Chatterjee}, \bibinfo{person}{Eric Knorr}, \bibinfo{person}{Michael
  Mitzenmacher}, \bibinfo{person}{Stratos Idreos}, {and} \bibinfo{person}{Tim
  Kraska}.} \bibinfo{year}{2022}\natexlab{}.
\newblock \showarticletitle{SNARF: a Learning-Enhanced Range Filter}.
\newblock \bibinfo{journal}{\emph{PVLDB}} (\bibinfo{year}{2022}).
\newblock


\bibitem[\protect\citeauthoryear{Wang, Dai, Chen, Li, Gu, Chai, Zheng, Chen,
  Li, Deng, et~al\mbox{.}}{Wang et~al\mbox{.}}{2024a}]%
        {Wang2024}
\bibfield{author}{\bibinfo{person}{Hancheng Wang}, \bibinfo{person}{Haipeng
  Dai}, \bibinfo{person}{Shusen Chen}, \bibinfo{person}{Meng Li},
  \bibinfo{person}{Rong Gu}, \bibinfo{person}{Huayi Chai},
  \bibinfo{person}{Jiaqi Zheng}, \bibinfo{person}{Zhiyuan Chen},
  \bibinfo{person}{Shuaituan Li}, \bibinfo{person}{Xianjun Deng},
  {et~al\mbox{.}}} \bibinfo{year}{2024}\natexlab{a}.
\newblock \showarticletitle{Bamboo Filters: Make Resizing Smooth and Adaptive}.
\newblock \bibinfo{journal}{\emph{IEEE/ACM Trans. Netw.}}
  (\bibinfo{year}{2024}).
\newblock


\bibitem[\protect\citeauthoryear{Wang, Dai, Li, Yu, Gu, Zheng, and Chen}{Wang
  et~al\mbox{.}}{2022}]%
        {Wang2022}
\bibfield{author}{\bibinfo{person}{Hancheng Wang}, \bibinfo{person}{Haipeng
  Dai}, \bibinfo{person}{Meng Li}, \bibinfo{person}{Jun Yu},
  \bibinfo{person}{Rong Gu}, \bibinfo{person}{Jiaqi Zheng}, {and}
  \bibinfo{person}{Guihai Chen}.} \bibinfo{year}{2022}\natexlab{}.
\newblock \showarticletitle{Bamboo Filters: Make Resizing Smooth}. In
  \bibinfo{booktitle}{\emph{ICDE}}.
\newblock


\bibitem[\protect\citeauthoryear{Wang, Guo, Yang, and Zhang}{Wang
  et~al\mbox{.}}{2024b}]%
        {Wang2024grf}
\bibfield{author}{\bibinfo{person}{Hengrui Wang}, \bibinfo{person}{Te Guo},
  \bibinfo{person}{Junzhao Yang}, {and} \bibinfo{person}{Huanchen Zhang}.}
  \bibinfo{year}{2024}\natexlab{b}.
\newblock \showarticletitle{GRF: A Global Range Filter for LSM-Trees with Shape
  Encoding}.
\newblock \bibinfo{journal}{\emph{Proceedings of the ACM on Management of
  Data}} (\bibinfo{year}{2024}).
\newblock


\bibitem[\protect\citeauthoryear{Wang, Zhou, Shi, and Qian}{Wang
  et~al\mbox{.}}{2019}]%
        {Wang2019}
\bibfield{author}{\bibinfo{person}{Minmei Wang}, \bibinfo{person}{Mingxun
  Zhou}, \bibinfo{person}{Shouqian Shi}, {and} \bibinfo{person}{Chen Qian}.}
  \bibinfo{year}{2019}\natexlab{}.
\newblock \showarticletitle{Vacuum Filters: More Space-Efficient and Faster
  Replacement for Bloom and Cuckoo Filters}.
\newblock \bibinfo{journal}{\emph{PVLDB}} (\bibinfo{year}{2019}).
\newblock


\bibitem[\protect\citeauthoryear{Williger and Maier}{Williger and
  Maier}{2019}]%
        {Williger2019}
\bibfield{author}{\bibinfo{person}{Robert Williger} {and}
  \bibinfo{person}{Tobias Maier}.} \bibinfo{year}{2019}\natexlab{}.
\newblock \emph{\bibinfo{title}{Concurrent Dynamic Quotient Filters: Packing
  Fingerprints into Atomics}}.
\newblock \bibinfo{thesistype}{Ph.D. Dissertation}. \bibinfo{school}{Karlsruher
  Institut f{\"u}r Technologie (KIT)}.
\newblock


\bibitem[\protect\citeauthoryear{Wu, He, Yan, Wu, Yang, Ruas, Zhang, and
  Cui}{Wu et~al\mbox{.}}{2021}]%
        {Wu2021}
\bibfield{author}{\bibinfo{person}{Yuhan Wu}, \bibinfo{person}{Jintao He},
  \bibinfo{person}{Shen Yan}, \bibinfo{person}{Jianyu Wu},
  \bibinfo{person}{Tong Yang}, \bibinfo{person}{Olivier Ruas},
  \bibinfo{person}{Gong Zhang}, {and} \bibinfo{person}{Bin Cui}.}
  \bibinfo{year}{2021}\natexlab{}.
\newblock \showarticletitle{Elastic Bloom Filter: Deletable and Expandable
  Filter Using Elastic Fingerprints}.
\newblock \bibinfo{journal}{\emph{IEEE Trans Comput}} (\bibinfo{year}{2021}).
\newblock


\bibitem[\protect\citeauthoryear{Xie, Min, Zhang, Wen, and Xie}{Xie
  et~al\mbox{.}}{2007}]%
        {Xie2007}
\bibfield{author}{\bibinfo{person}{Kun Xie}, \bibinfo{person}{Yinghua Min},
  \bibinfo{person}{Dafang Zhang}, \bibinfo{person}{Jigang Wen}, {and}
  \bibinfo{person}{Gaogang Xie}.} \bibinfo{year}{2007}\natexlab{}.
\newblock \showarticletitle{A Scalable Bloom Filter for Membership Queries}.
\newblock \bibinfo{journal}{\emph{GLOBECOM}} (\bibinfo{year}{2007}).
\newblock


\bibitem[\protect\citeauthoryear{Xie, Chen, Wang, Wang, Tao, and Cheng}{Xie
  et~al\mbox{.}}{2022}]%
        {Wie2022}
\bibfield{author}{\bibinfo{person}{Minghao Xie}, \bibinfo{person}{Quan Chen},
  \bibinfo{person}{Tao Wang}, \bibinfo{person}{Feng Wang},
  \bibinfo{person}{Yongchao Tao}, {and} \bibinfo{person}{Lianglun Cheng}.}
  \bibinfo{year}{2022}\natexlab{}.
\newblock \showarticletitle{Towards Capacity-Adjustable and Scalable Quotient
  Filter Design for Packet Classification in Software-Defined Networks}.
\newblock \bibinfo{journal}{\emph{IEEE Open Journal of the Computer Society}}
  (\bibinfo{year}{2022}).
\newblock


\bibitem[\protect\citeauthoryear{Zhang, Chen, Jin, and Reviriego}{Zhang
  et~al\mbox{.}}{2021}]%
        {Zhang2021}
\bibfield{author}{\bibinfo{person}{Fan Zhang}, \bibinfo{person}{Hanhua Chen},
  \bibinfo{person}{Hai Jin}, {and} \bibinfo{person}{Pedro Reviriego}.}
  \bibinfo{year}{2021}\natexlab{}.
\newblock \showarticletitle{The Logarithmic Dynamic Cuckoo Filter}.
\newblock \bibinfo{journal}{\emph{ICDE}} (\bibinfo{year}{2021}).
\newblock


\bibitem[\protect\citeauthoryear{Zhang, Lim, Leis, Andersen, Kaminsky, Keeton,
  and Pavlo}{Zhang et~al\mbox{.}}{2018}]%
        {Zhang2018a}
\bibfield{author}{\bibinfo{person}{Huanchen Zhang}, \bibinfo{person}{Hyeontaek
  Lim}, \bibinfo{person}{Viktor Leis}, \bibinfo{person}{David~G Andersen},
  \bibinfo{person}{Michael Kaminsky}, \bibinfo{person}{Kimberly Keeton}, {and}
  \bibinfo{person}{Andrew Pavlo}.} \bibinfo{year}{2018}\natexlab{}.
\newblock \showarticletitle{{SuRF: Practical Range Query Filtering with Fast
  Succinct Tries}}.
\newblock \bibinfo{journal}{\emph{SIGMOD}} (\bibinfo{year}{2018}).
\newblock


\end{thebibliography}
